\documentclass[twocolumn,prb,aps,amsmath,showpacs]{revtex4-1}
\usepackage{graphics}
\usepackage{graphicx}
\usepackage{epsfig}
\usepackage{subfigure}
\usepackage[dvips]{color}
\usepackage{amssymb}

\begin{document}

\title{Electronic Structure, Transport and Phonons of SrAg$Ch$F ($Ch$=S, Se, Te): Bulk Superlattice Thermoelectrics}
\author{Vijay Kumar Gudelli and V. Kanchana$^*$}
\affiliation{Department of Physics, Indian Institute of Technology Hyderabad, Ordnance Factory Estate, Yeddumailaram-502 205, Telangana, India}
\author{G. Vaitheeswaran$^*$}
\affiliation{Advanced Centre of Research in High Energy Materials (ACRHEM), University of Hyderabad, Prof. C. R. Rao Road, Gachibowli, Hyderabad-500 046, Telangana, India}
\author{David J. Singh}
\affiliation{Materials Science and Technology Division, Oak Ridge National Laboratory, Oak Ridge, TN 37831, USA}
\author{A. Svane and N. E. Christensen}
\affiliation{Department of Physics and Astronomy, Aarhus University, DK-8000 Aarhus C, Denmark}
\author{S. D. Mahanti}
\affiliation{Department of Physics and Astronomy, Michigan State University, East Lansing, Michigan 48824, USA}

\begin{abstract}
We report calculations of the electronic structure, vibrational properties and transport for the p-type semiconductors, SrAg$Ch$F ($Ch$=S, Se and Te). We find soft phonons with low frequency optical branches intersecting the acoustic modes below 50 $cm^{-1}$, indicative of a material with low thermal conductivity. The bands at and near the valence band maxima are highly two dimensional, which leads to high thermopowers even at high carrier concentrations, which is a combination that suggests good thermoelectric performance. These materials may be regarded as bulk realizations of superlattice thermoelectrics.
\end{abstract}

\maketitle

\section{Introduction}
Thermoelectrics enable direct solid state conversion of thermal and electrical energy.\cite{Ioffe,Wood} They are widely used in spacecraft power and in terrestrial cooling and may have wide applications in energy technologies such as waste heat recovery if their efficiencies can be improved.\cite{Yang} The efficiency that can be achieved is limited by the properties of available thermoelectric materials, and in particular a dimensionless, but temperature dependent figure of merit $ZT = \sigma S^2T/\kappa$, where $\sigma$ is electrical conductivity, $T$ is absolute temperature, $S$ is the Seebeck coefficient, also known as the thermopower, and $\kappa$ is the thermal conductivity. Besides applications, finding high ZT is an interesting fundamental problem because it requires obtaining a contraindicated combination of transport properties, e.g. high electrical conductivity with low thermal conductivity, and high thermopower with high conductivity.  Approaches that have been used effectively used to discover new high ZT thermoelectrics include approaches focusing on thermal conductivity and those emphasizing electronic transport. Thermal conductivity has been addressed via phonon engineering, e.g. via rattling ions \cite{XShi} and nano-structuring \cite{Delaire} among other approaches. Electronic transport has involved complex dopants,\cite{Poudel} interfaces,\cite{HLiu} dimensional reduction in nanostructured materials\cite{Hicks} and unusual electronic structures.\cite{H.Shi,Parker1}


Dimensional reduction is among the most influential concepts in thermoelectrics. This originated with Hicks and Dresselhaus \cite{Hicks}. They showed that the electronic properties could be enhanced by putting thermoelectric materials in superlattice structures with insulating barrier layers and that the improvement is largest for the shortest period superlattices. This stimulated a large body of subsequent work on nanostructured thermoelectrics including work utilizing interface phonon scattering to increase the figure of merit \cite{Chen,Hochbaum} and showing that low dimensional electronic features leading to enhanced performance can occur in three dimensional materials as well \cite{Parker1}. Nonetheless, it has proved challenging to use these ideas to develop practical high performance thermoelectric materials. This is partly because of the obvious difficulty of using superlattices in a bulk application, but also because of more fundamental reasons. Barrier layer materials that can be readily grown in superlattices with thermoelectrics tend to have modest band gaps, and therefore the barrier layers must be relatively thick before the dimensional reduction is effective. However, while low thermal conductivity is important for thermoelectric performance, both the barrier and active layers will contribute to the heat conduction.
\par
An alternate approach may be the use of crystals that naturally have the features of superlattices, such as near two dimensional electronic structures. In this regard, it is noteworthy that the best performing oxide thermoelectric is p-type Na$_x$CoO$_2$ \cite{Terasaki}, which has a very two dimensional electronic structure \cite{DJS1}. Hosono and co-workers \cite{Ueda} have pioneered the use of charged layers in crystal chemistry to design layered compounds for transparent conductors and high temperature superconductors \cite{Kamihara}. This approach, which is a generalization of the Zintl chemistry, consists of building compounds out of charge compensating layers, specifically a stable cationic layer, e.g. SrF or LaO, and an anionic functional layer, e.g. FeAs, to make compounds like LaFeAsO \cite{Kamihara} or SrFeAsF \cite{Han}. Key to this is the strong affinity of atoms in the cationic layer (La-O, Sr-F) for each other, and the fact that the anionic and cationic layers are bound into the structure by Coulomb interactions. This typically yields remarkably stable materials, with good mechanical properties. On the other hand, this strong cohesion can also lead to three dimensional electronic properties even though structurally the compounds are layered.
\par
Returning to thermoelectrics, this chemistry potentially provides a scheme for making bulk layered thermoelectric materials separated by thin strongly insulating barrier layers, while maintaining chemical stability. Furthermore, with judicious choice of barriers, it may be possible to design compounds with strong phonon scattering. Here we present electronic structure and transport calculations for SrAg$Ch$F, $Ch$=S,Se,Te which are examples of such compounds. They consist of anionic Ag$Ch$ layers, related to the thermoelectrics Ag$_2$Se and Ag$_2$Te \cite{Ferhat,Mi,Taylor,Pei} and Ag containing chalcopyrites, such as AgGaTe$_2$ \cite{Yusufu,Parker2}, with intervening SrF layers. We find that indeed the materials are electronically rather two dimensional in their valence band structures in accord with prior work by Banikov and co-workers \cite{Bannikov} and that this is reflected in the transport, effectively a natural superlattice of Ag$Ch$ thermoelectric. This two dimensionally results in very favorable electronic properties for a thermoelectric.

\section{Methodology}
We did transport calculations based on the electronic structures as calculated using the full-potential linearized augmented plane wave (FLAPW) method as implemented in the WIEN2k code \cite{Blaha}. The crystal structure of SrAg$Ch$F ($Ch$=S,Se,Te) is tetragonal with the space group P4/nmm \cite{Charkin}, and may be viewed as alternating blocks of [SrF] and [Ag$Ch$] as shown in Fig. 1(a). The structure consists of four inequivalent atomic positions Sr at $2c$ site (1/4, 1/4, Z$_{Sr}$), F at $2a$ site (3/4, 1/4, 0), Ag at $2b$ site (3/4, 1/4, 1/2) and chalcogen at $2c$ site (1/4,1/4 Z$_{Ch}$), where Z$_{Sr}$ and Z$_{Ch}$ are the internal co-ordinates of Sr and the chalcogen, respectively. We optimized the structures using the generalized gradient approximation of Perdew, Burke and Ernzerhof (PBE) \cite{PBE}. The resulting structures are given in Table II. We then did electronic structure calculations using these optimized structures. For this purpose we used the Tran-Blaha modified Becke-Johnson potential (TB-mBJ) \cite{Becke}. This functional yields improved band gaps compared to the PBE and related functionals \cite{Koller,DJS2}. This can be important for an accurate description of electronic transport. These calculations were done self-consistently and included spin-orbit interaction (SOC). The transport properties of SrAg$Ch$F were obtained using semi-classical Boltzmann transport theory as implemented in the BOLTZTRAP code \cite{GKH} with rigid bands and the constant scattering time ($\tau$) approximation. The TB-mBJ potential functional is optimized to reproduce band gaps of semiconductors and insulators and cannot be used for energies and structural properties. For the phonon and elastic properties we used the PBE-GGA, which is a well-tested energy functional. Phonon dispersion calculations were performed within the framework of density functional perturbation theory (DFPT) as implemented in the plane wave self-consistent field (PW-scf) \cite{pwscf}. In addition, we have computed the elastic properties and analysed the nature of bonding using the plane wave pseudopotential method as implemented in CASTEP \cite{Payne,Segall}. We cross-checked the results by comparing with elastic constants obtained using WIEN2k and obtained close agreement.

\section{Results and discussion}
\subsection{Electronic band structure}
We start with the electronic structure. The calculated band structures for all the investigated compounds are presented in Fig. 2. The corresponding densities of states (DOS) are given in Fig. 2(d). The general band shapes of all the compounds are similar. The Sr-F layers are clearly ionic (see Table IV). As may be seen in the density of states, the F $2p$ bands are at high binding energy, several $eV$ below the valence band maximum (VBM). The VBM comes from hybridized Ag-d and Ch-p states. This makes the valence bands relatively heavy. The calculated band masses are given in Table III. This suggests the possibility of good thermoelectric performance in analogy with other semiconductors with monovalent Ag. However, the details of the transport depend on the band dispersions. We note that the effect of SOC on the bands is not negligible for these compounds. We find a substantial SOC induced lifting of the degeneracy of the bands at the $\Gamma$ point as compared to the results Bannikov et. al, \cite{Bannikov}. The SOC splitting is found to be 31 $meV$ for S, 129 $meV$ for Se and 330 $meV$ for Te. These are sizable on the scale of $kT$, for temperatures of interest and are also a little higher than found in the Cu-based 1111 family, BaCu$Ch$F \cite{Zakutayev}. The VBM and the conduction band minimum (CBM) are both located at the centre of the Brillouin zone, i.e. $\Gamma$-point, making the investigated compounds direct-band-gap semiconductors. The band gap values are found to vary non-monotonously from S to Te.  An interesting feature to note is the presence of quasi-flat valence bands along the $\Gamma$-$Z$, $R$-$X$, $M$-$A$ directions of the Brillouin zone along with the highly dispersive bands in the other high symmetry directions. Thus from an electronic point of view these are very two dimensional compounds. Importantly, the two top valence bands in all three compounds are extremely flat along the $\Gamma$-$Z$ direction. This is also reflected in the effective masses given in the Table III and also in the charge density plots shown in Fig. 1(b), where the intra layer bonding in the Ag-chalcogen block is of mixed ionic and covalent nature, while the inter layer bonding between the SrF and Ag$Ch$ blocks is predominantly ionic (see Table IV). It is this ionic character that leads to the effectively two dimensional electronic structure. In particular, the Sr and F atoms do not contribute to states near the valence band maximum, and therefore hopping through orbitals on these atoms is suppressed and the c-axis dispersion of the upper valence bands is very small. On the other hand, the conduction bands show significant dispersion for k-vector variation along the tetragonal axis, i.e. the virtues of two-dimensionality with respect to thermoelectric performance are likely more dominant for p-type doping than for n-type doping.

\subsection{Thermoelectric properties}
Turning to the thermoelectric properties, we present data up to 900 $K$ in the carrier concentration range $1\times 10^{18}$ to $1\times 10^{21} cm^{-3}$ for both electrons and holes. We note that in analogy with other monovalent Ag compounds, it is clear that the material should form good p-type thermoelectric. Therefore we focus the discussion on p-type although for comparison we give results for n-type. The main results for the electronic transport of SrAgSF are given in Fig. 3. which show the n- and p-type thermopowers for SrAgSF (the electronic transport results see Fig. 6 and Fig. 7 for SrAgSeF and SrAgTeF respectively). The figures show the in plane and c-axis thermopowers, $S$, and transport functions $\sigma/\tau$ ($\sigma$ is the conductivity, and $\tau$ is the inverse scattering rate). First of all, there is a very strong anisotropy of $\sim 10^2$ in $\sigma/\tau$, and therefore also $\sigma$, for p-type transport, with a smaller anisotropy for n-type. Thus, particularly for p-type these are highly two dimensional semiconductors. Secondly, there is also a substantial anisotropy in $S$. This is unusual in three dimensional semiconductors, even when they are anisotropic. This behavior can arise in anisotropic semiconductors due to bipolar effects \cite{Parker3}, but due to the sizable band gaps this cannot be the case here. Alternately two dimensionality of the electronic structure, leading to open Fermi surfaces when doped can lead to this type of behavior \cite{Ong,Magnuson}. This is the case here. Thus both the conductivity and thermopower indicate practically two dimensional transport for p-type samples.

For these two dimensional p-type materials, the conductivity in the c-axis direction will be very low. This means that thermoelectric performance of crystals can only be good in the in-plane direction. The thermopower of a composite is generally a weighted average of the thermopowers in different directions. When the electrical conductivity anisotropy is much stronger than that of the thermal conductivity, this weighting is dominated by the electrical conductivity, and in a material as anisotropic as these p-type materials, the measured thermopower of a ceramic sample will be determined by the in-plane value. The measured electrical conductivity of a ceramic on the other hand will be reduced relative to the in-plane direction of a crystal both by grain boundary resistance and by the resistance of grains with the ab-plane oriented away from the direction of the average current. Since the thermoelectric figure of merit is $ZT=\sigma S^2 T/\kappa$, it is clear then that the thermoelectric performance of a ceramic of these two dimensional materials will be poorer than that of in-plane oriented crystals, and that the performance will be best for a highly textured ceramic. In the following we focus on the in-plane transport for crystalline samples, which is also the limiting behaviour for highly c-axis textured ceramic.

As discussed by Hicks and Dresselhaus \cite{Hicks}, reduction in the dimensionality of a semiconductor improves the electronic contribution to $ZT$. The density of states near the band edge for a 2D material has a step function shape, and is therefore higher than for a 3D semiconductor with a similar effective mass. This leads to an enhanced thermopower for a given carrier concentration. In the low temperature limit, $S(T)\propto T/E_F$ where $E_F$ is the Fermi energy relative to the band edge, and therefore since higher density of states means lower $E_F$ for a given carrier concentration, dimensional reduction enhances the thermopower. Equivalently, one can consider a fixed Fermi level, $E_F$, i.e. doping to similar $S(T)$ for the 2D and 3D semiconductors. In that case, the 2D semiconductor will have a higher carrier density but the same in-plane Fermi velocity, and therefore enhanced conductivity relative to the 3D material.

Most good 3D thermoelectric materials have carrier concentrations of $\sim 10^{19} cm^{-3}$ and thermopower values for the peak $ZT$ in the range 200 - 300 $\mu V/K$. This can be rationalized from the formula $ZT = rS^2/L$, with $r=\kappa_e/\kappa$, where $\kappa_e$ is the electronic part of the thermal conductivity, written via the Wiedemann-Franz relation as $\kappa_e=L\sigma T$. With the standard value of the Lorentz number, $L=L_0$, and $r=1$, $ZT=1$ corresponds to $S\sim 160  \mu V/K$. When the electronic thermal conductivity dominates, $r\sim 1$ and $ZT$ can be increased by increasing $S$, i.e. lowering the carrier concentration. On the other hand, when lattice thermal conductivity is dominant, $r\propto \kappa_e\propto \sigma$,  and $ZT$ is increased by increasing the power factor, $\sigma S^2$. This can usually be accomplished through higher doping levels without decreasing $S$. Thus values of $r$ of very roughly 0.5 are common in highly optimized thermoelectric materials, which with the state-of-the-art values of $ZT$ in the range 1 - 2, leads to the above expected range for $S$.

The p-type thermopower data of Fig. 3 (and in Figs. 6-7), clearly show the effect of the 2D electronic structure. In particular, one observes very high values of the thermopower at high carrier concentrations, even at 300 $K$. For SrAgSF and SrAgSeF, in-plane thermopowers of 200 $\mu V/K$ are obtained at 300 $K$ for carrier concentrations up to $10^{20} cm^{-3}$ and higher, while for the telluride such values occur up to more than $3\times 10^{19} cm^{-3}$. Due to the relatively large band gaps of these compounds, the in-plane $S(T)$ increase monotonically with $T$. Although in general the conductivity would be expected to decrease with $T$, it is likely that the increase in $S$ will more than offset this for optimized thermoelectric performance and that the peak $ZT$ will increase with $T$ in these compounds up to the maximum temperature for which they are stable. This is a consequence of the sizable band gaps. For all three compounds we find p-type, in-plane $S(T)$ well above 200 $\mu V/K$ at a carrier concentration of $10^{20}$ holes  $cm^{-3}$. Electronic transport results at 700 $K$ are summarized in Table I. This indicates that one can anticipate good thermoelectric performance in these compounds provided that the lattice thermal conductivity is not too high.

In order to better define the carrier concentration range, we plot the transport function $S^2 \sigma/\tau$ in Fig. 4. This function is related to the power factor. In general the inverse scattering rate, $\tau$, decreases with temperature and doping level. Therefore the position of the peak in this transport function provides an upper bound on the possible optimum doping level. As seen, this is $\sim 10^{21}$ holes  $cm^{-3}$ for all three compounds. SrAgSF and SrAgSeF give higher values of this transport function than does SrAgTeF. This is a consequence of the lower $S(T)$ in the telluride. It is possible, however, that this could be compensated by a difference in mobility, i.e. if the mobility at a given carrier concentration is higher in the telluride, or by a lower lattice thermal conductivity. In any case from the electronic point of view all three compounds are highly two dimensional for p-type doping and show promising thermoelectric properties as a consequence.

\subsection{Lattice Dynamics}
We now turn to the lattice dynamical properties. The single crystal elastic constants were calculated and are given in Table V, along with polycrystalline averages that may be useful for comparison with future experiments on ceramic samples. The compounds are found to be mechanically stable as expected and the computed Debye temperatures of the SrAg$Ch$F compounds are similar to that of BiCuSO \cite{Liu}. To be more precise, SrAgSeF has the lowest value of the Debye temperature which might indicate a lower thermal conductivity. We give the phonon dispersions and density of states of SrAgSeF in Figs. 5(a),(b). We find low frequency optic branches that intersect the acoustic branches starting well below 50 $cm^{-1}$. 
This suggests strong phonon-phonon scattering. Thermal conductivity in bulk semiconductors at temperatures relevant to thermoelectrics is normally intrinsically limited by anharmonic phonon-phonon scattering\cite{Ziman} and therefore this suggests low thermal conductivity. From Fig 5(b), it is clearly evident that these lower frequency modes originate mainly due to the AgSe-inter layer vibrations. We have also plotted the lower acoustic and optic modes separately in Fig. 5(c). A similar behaviour is also observed in BiCuSeO, where the experimentally reported thermal conductivity is low at high temperature \cite{Li}, and the phonon dispersion plots show a similar behaviour.

\section{Summary}
We have discussed the electronic structure, transport and vibrational properties of SrAg$Ch$F. We find that especially for p-type doping these are highly two dimensional materials, in effect natural superlattices. This dimensional reduction is reflected in the electronic transport properties relevant for thermoelectric performance, with high values of the thermopower at high doping levels. We also find phonon dispersions with low lying optic modes cutting the acoustic branches, indicative of materials with low thermal conductivity. These results suggest exploration of the thermoelectric properties of p-type SrAg$Ch$F materials for in-plane oriented crystals. It is expected that the optimum carrier concentrations will be in the $10^{20} cm^{-3}$ range for temperatures above ambient and that the optimized $ZT$ will increase with temperature up to temperatures close to the limit of stability of the compounds.

\section{Acknowledgement}
\textit{Acknowledgement} The auhtors V.K.G and V.K would like to acknowledge IIT-Hyderabad for providing computational facility. V.K.G. would like to thank MHRD for the fellowship. G.V thanks the Center for Modelling Simulation and Design-University of Hyderabad (CMSD-UoH) for providing computational facility. Work at ORNL was supported by the Department of Energy, Office of Science, Basic Energy Sciences, through the Solid-State Solar-Thermal Energy Conversion (S3TEC) Center.

*Authors for Correspondence, \\
E-mail: kanchana@iith.ac.in, gvsp@uohyd.ernet.in

\begin{table*}
\caption{Thermopower, electrical conductivity and power factor of SrAg$Ch$F for carrier concentrations of $10^{19}$ and $10^{20} cm^{-3}$ at 700 $K$.}
\begin{tabular}{cccccccccccccccccc}
\hline
		&concentration $(cm^{-3})$			&	&10$^{19}$	&	&	&	&10$^{20}$\\
\hline
		&carriers 					&n$_h$	&	&n$_e$	&	&n$_h$	&	&n$_e$	&	\\
\hline
		&Direction					&a	&c	&a	&c	&a	&c	&a	&c	\\
\hline
SrAgSF		&S($\mu V/K$)					&554	&662	&339	&317	&355	&465	&162	&130	\\
		&($\sigma/\tau)x10^{17}(\Omega ms)^{-1}$	&1.49	&0.03	&4.95	&3.05	&14.68	&0.28	&46.34	&26.25	\\
		&(S$^2\sigma/\tau)x10^{11}(W/mK^2s)$		&0.46	&0.01	&0.57	&0.31	&1.85	&0.06	&1.22	&0.45	\\		
\hline
SrAgSeF		&S($\mu V/K$)					&514	&650	&306	&297	&451	&319	&132	&119	\\
		&($\sigma/\tau)x10^{17}(\Omega ms)^{-1}$	&2.10	&0.02	&6.08	&4.38	&20.51	&0.23	&55.56	&38.19	\\
		&(S$^2\sigma/\tau)x10^{11}(W/mK^2s)$		&0.56	&0.01	&0.57	&0.39	&4.18	&0.024	&0.98	&0.54	\\		
\hline
SrAgTeF		&S($\mu V/K$)					&437	&593	&282	&281	&243	&396	&117	&111	\\
		&($\sigma/\tau)x10^{17}(\Omega ms)^{-1}$	&3.51	&0.009	&6.86	&5.38	&33.84	&0.10	&59.74	&46.34	\\
		&(S$^2\sigma/\tau)x10^{11}(W/mK^2s)$		&0.67	&0.003	&0.55	&0.43	&2.01	&0.02	&0.82	&0.58	\\		
\hline
\end{tabular}
\end{table*}

\begin{figure*}
\begin{center}
\subfigure[]{\includegraphics[width=60mm,height=60mm]{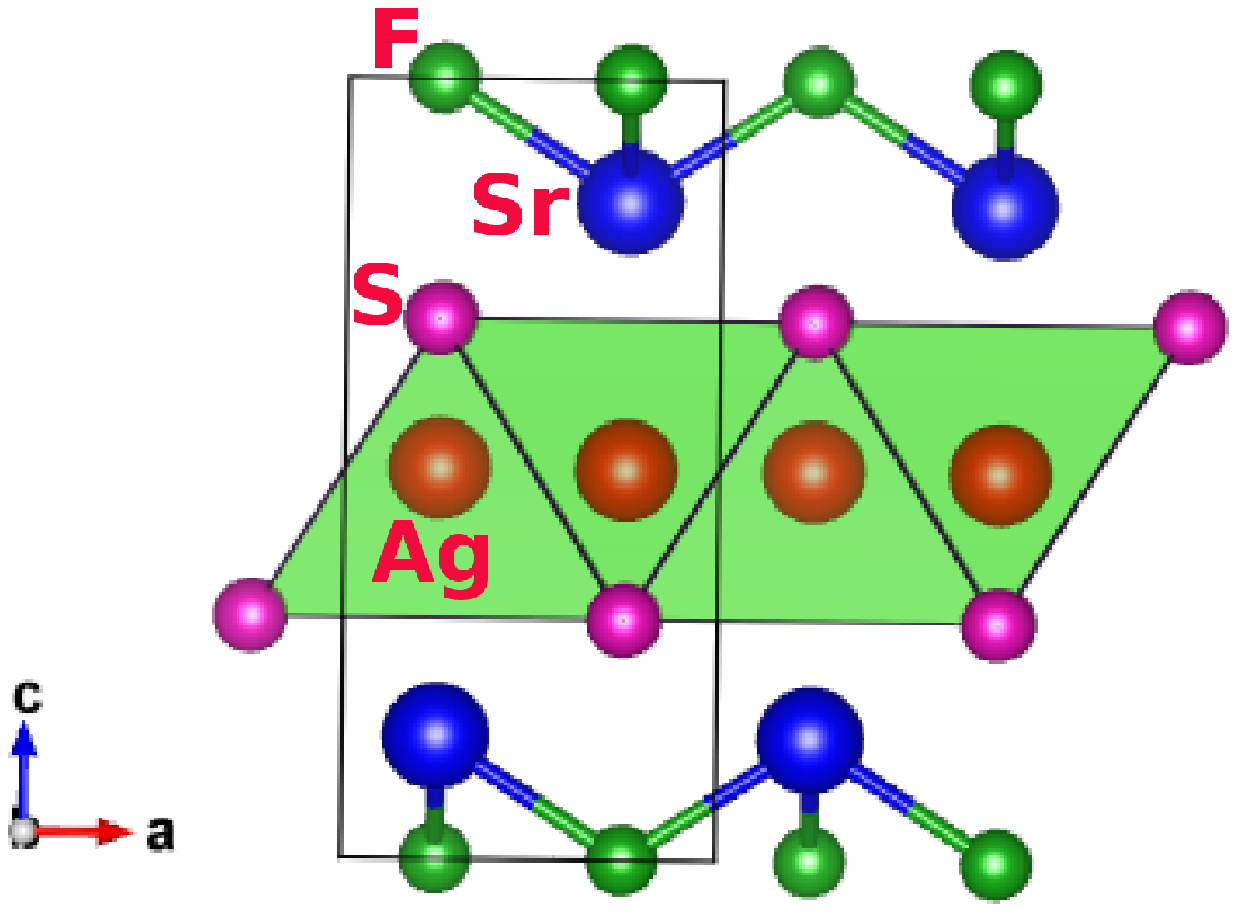}}
\subfigure[]{\includegraphics[width=75mm,height=70mm]{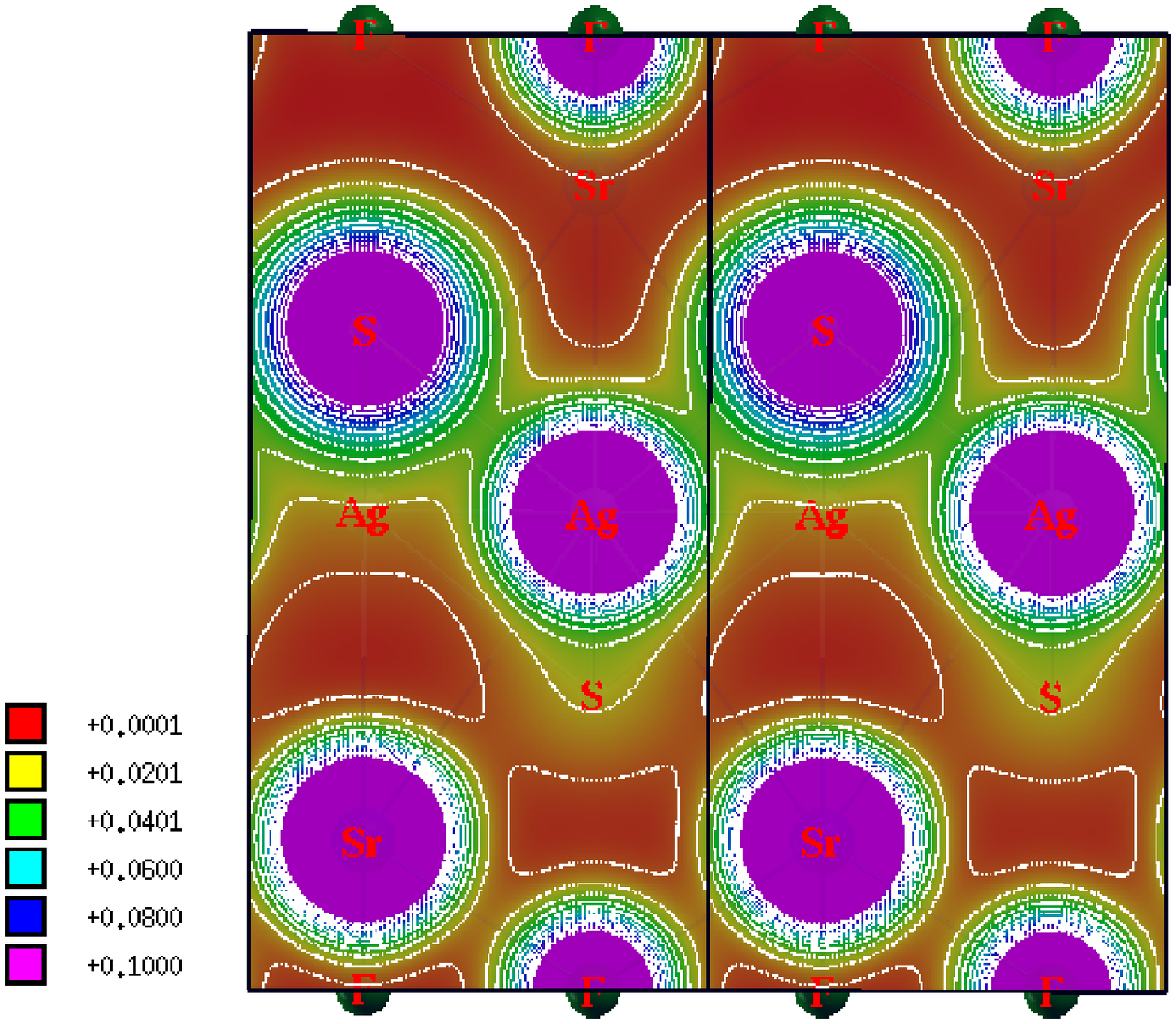}}
\caption{(Color online) (a) Crystal structure and (b) charge density of SrAgSF}
\end{center}
\end{figure*}

\begin{figure*}
\begin{center}
\subfigure[]{\includegraphics[width=55mm,height=55mm]{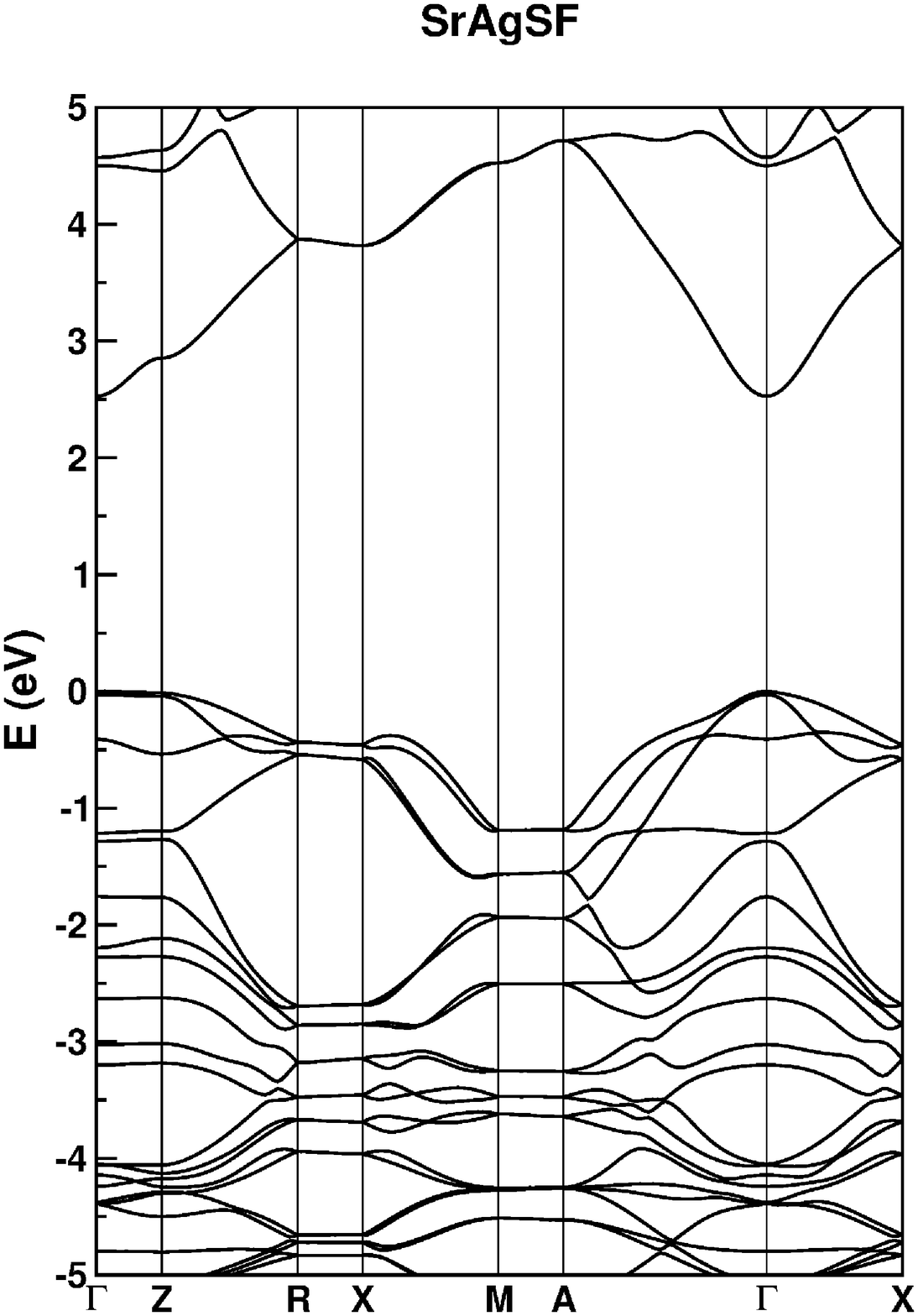}}
\subfigure[]{\includegraphics[width=55mm,height=55mm]{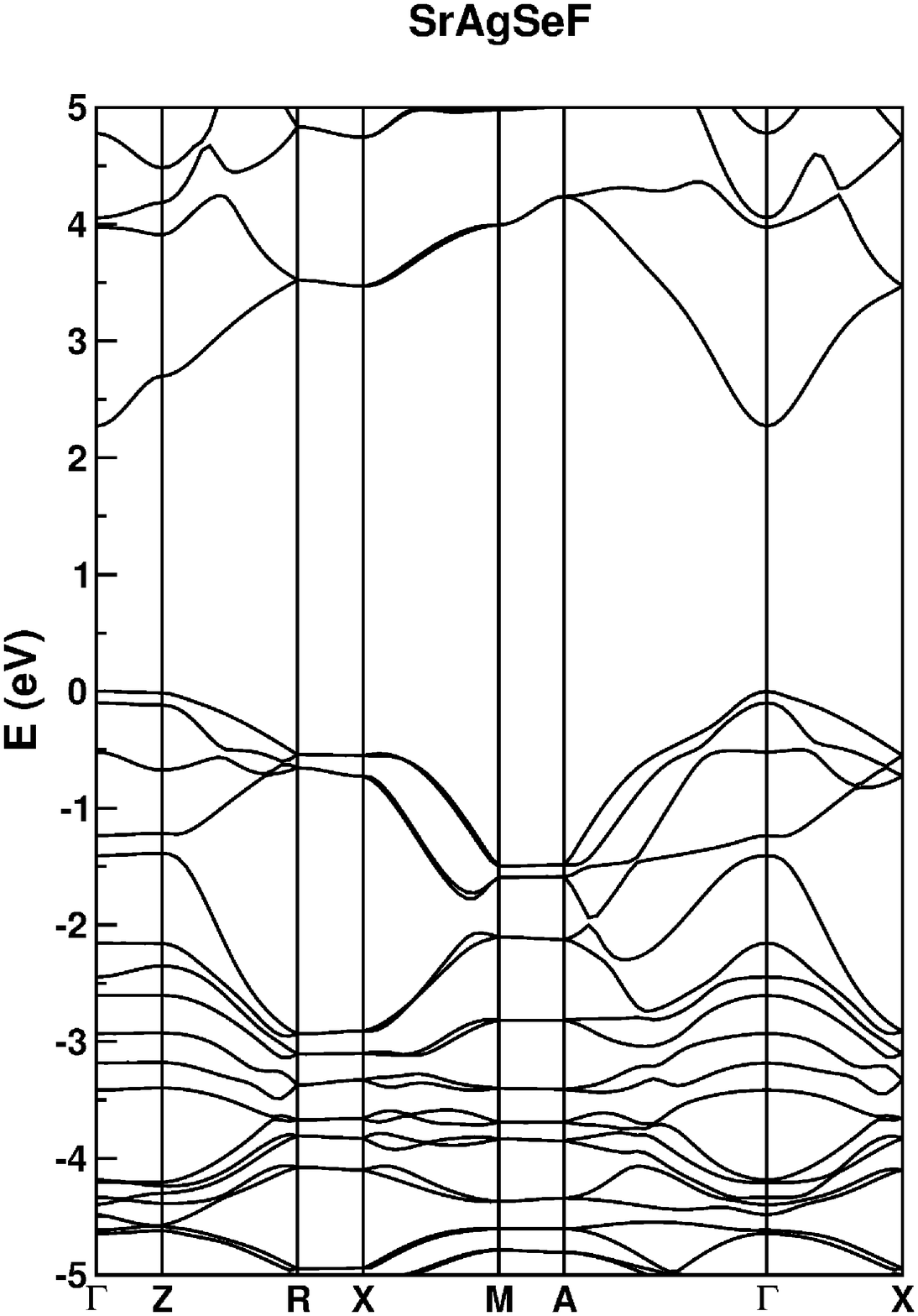}}
\subfigure[]{\includegraphics[width=55mm,height=55mm]{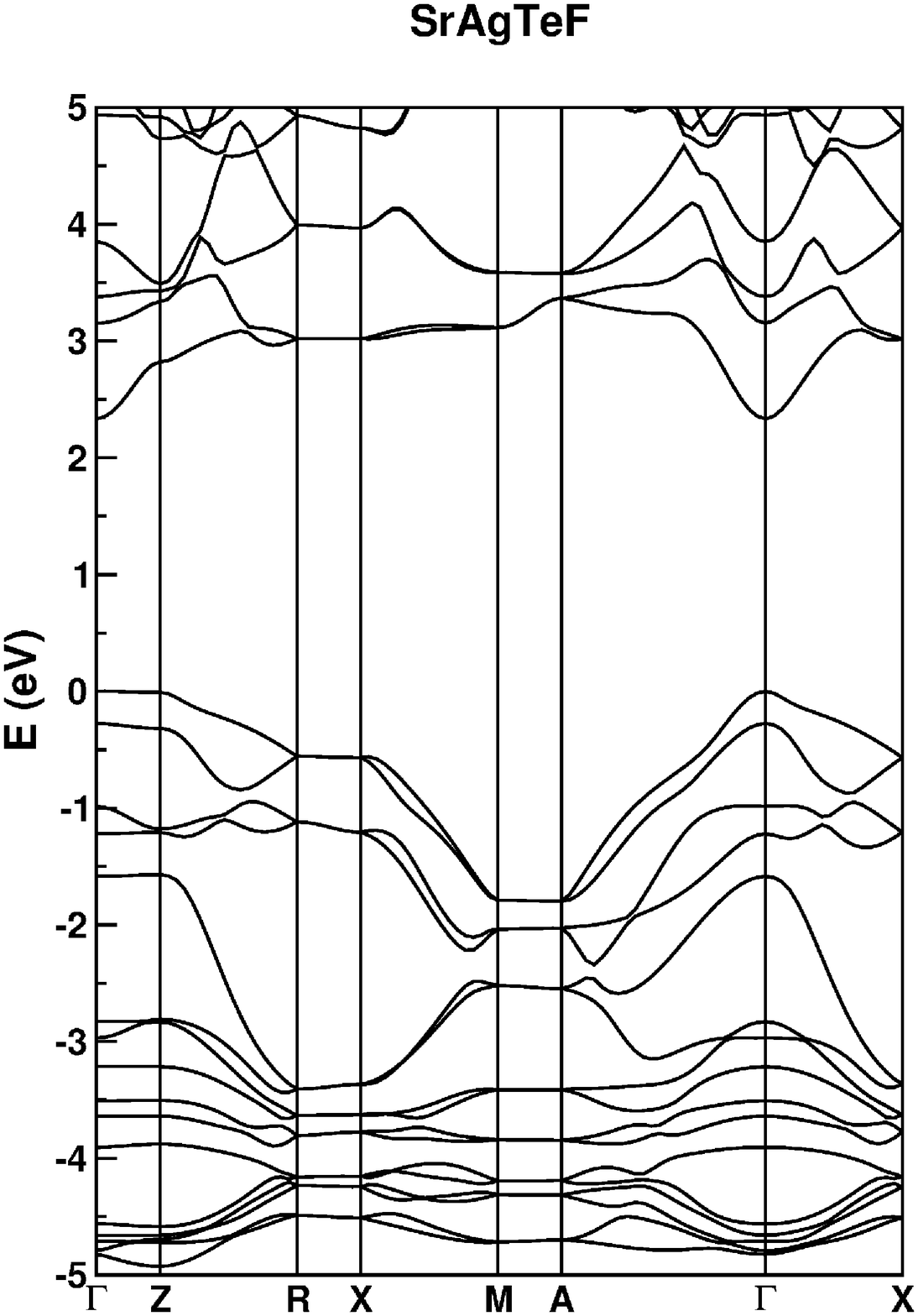}}\\
\subfigure[]{\includegraphics[width=60mm,height=60mm]{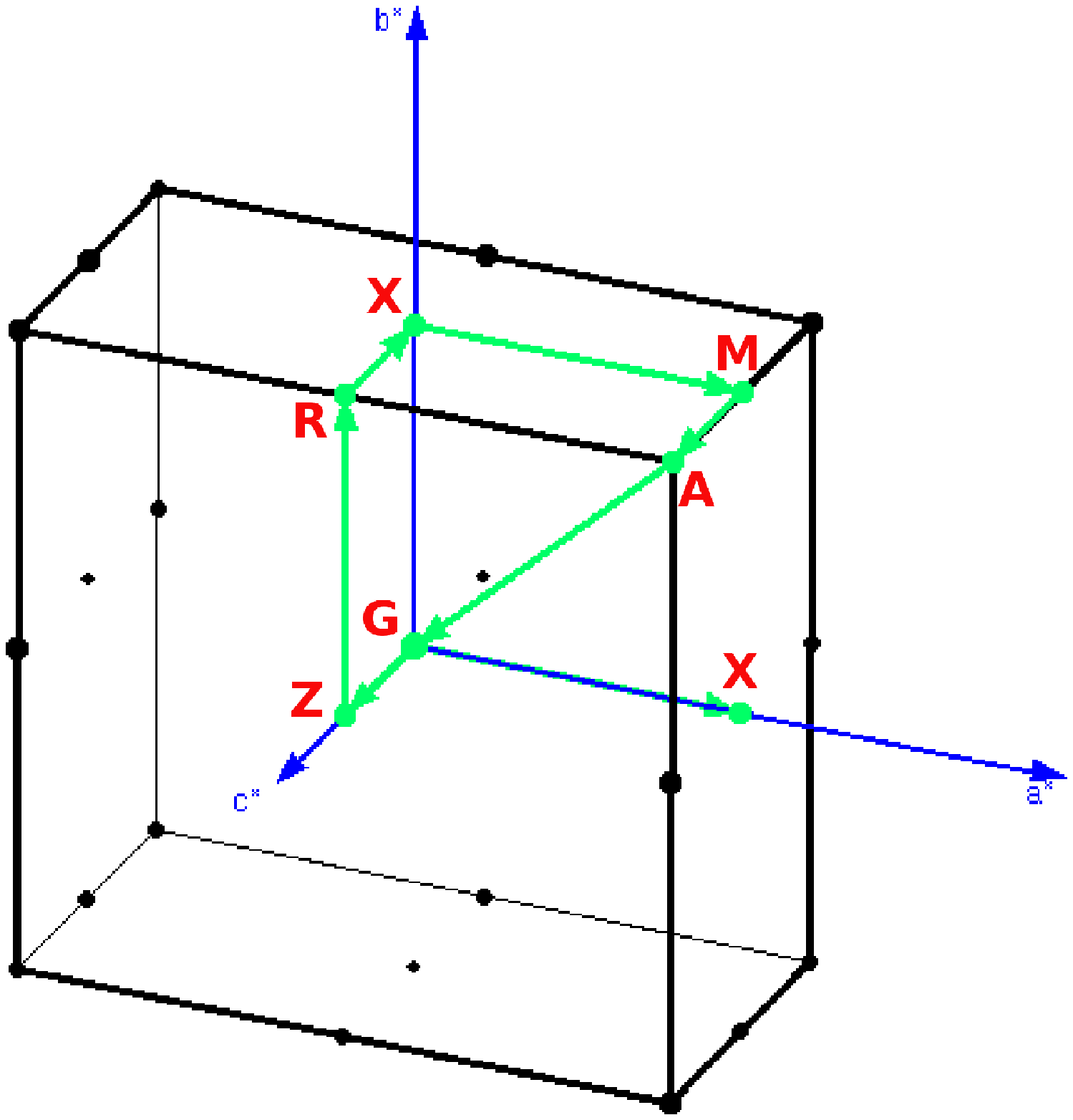}}\\
\subfigure[]{\includegraphics[width=100mm,height=80mm]{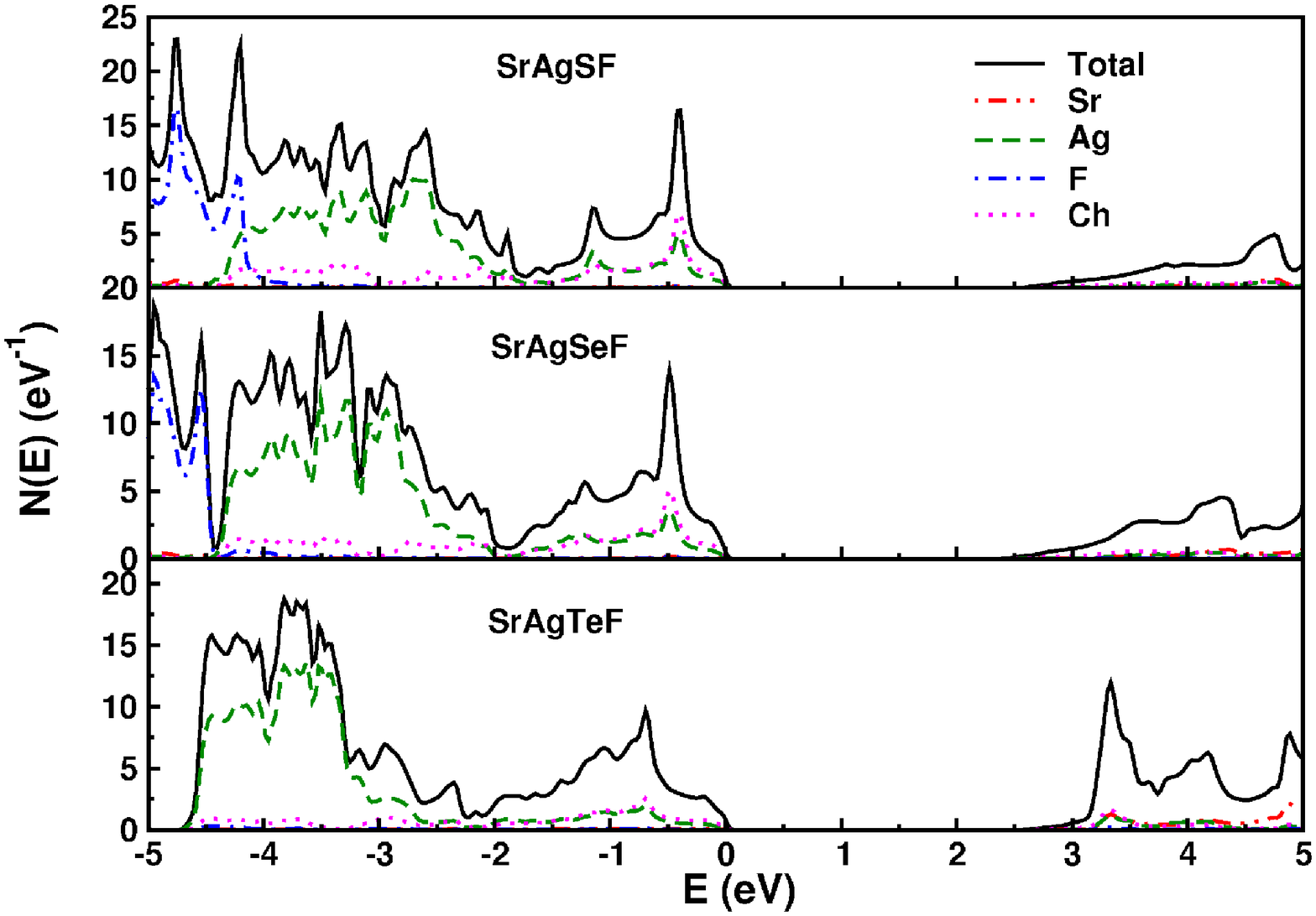}}
\caption{(Color online) The calculated band structures of (a) SrAgSF (b) SrAgSeF and (c) SrAgTeF. (d) shows the Brillouin zone path direction (e) shows the calculated density of states of SrAg$Ch$F ($Ch$ = S, Se and Te)}
\end{center}
\end{figure*}

\begin{figure*}
\begin{center}
\subfigure[]{\includegraphics[width=70mm,height=70mm]{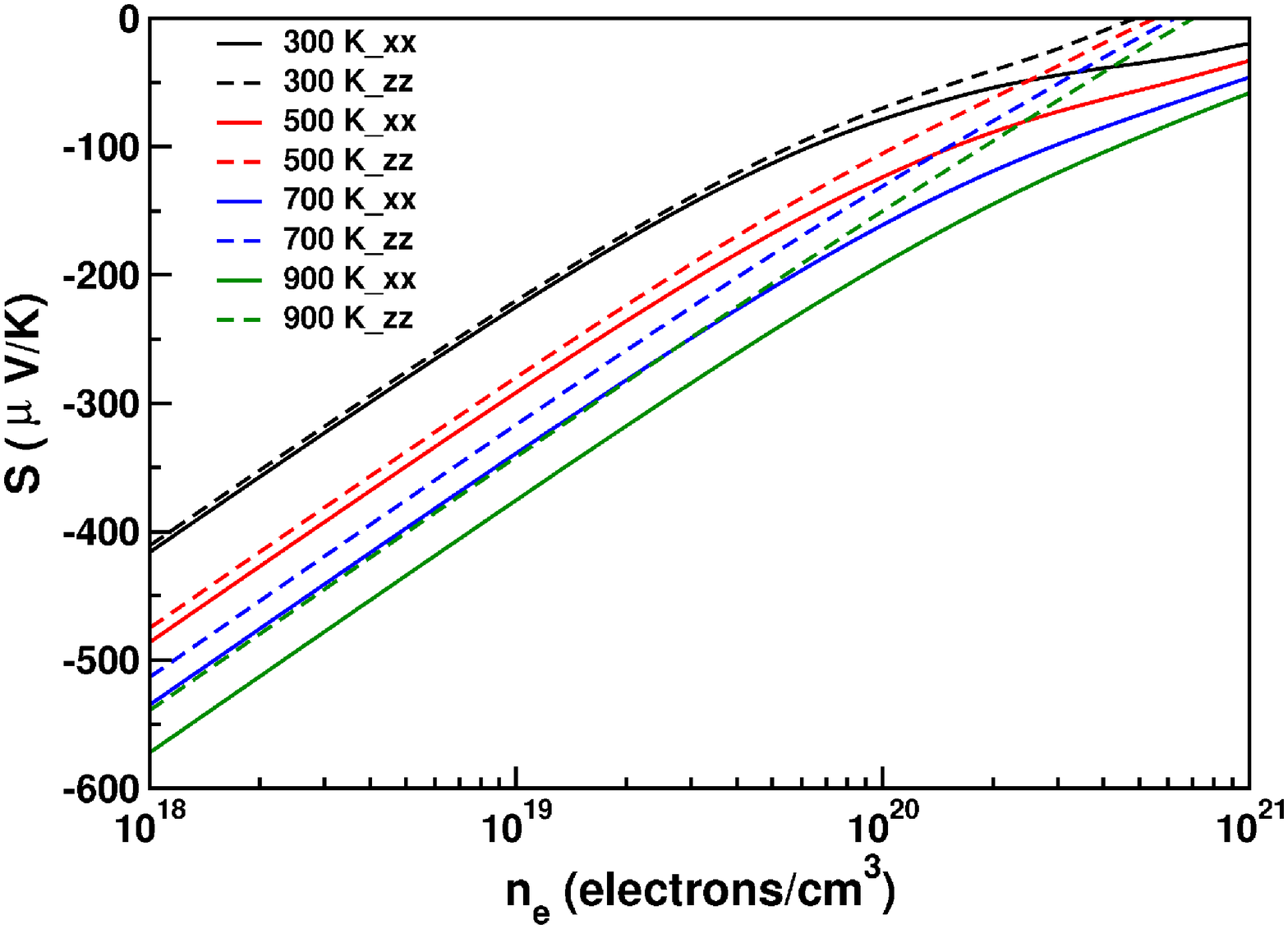}}
\subfigure[]{\includegraphics[width=70mm,height=70mm]{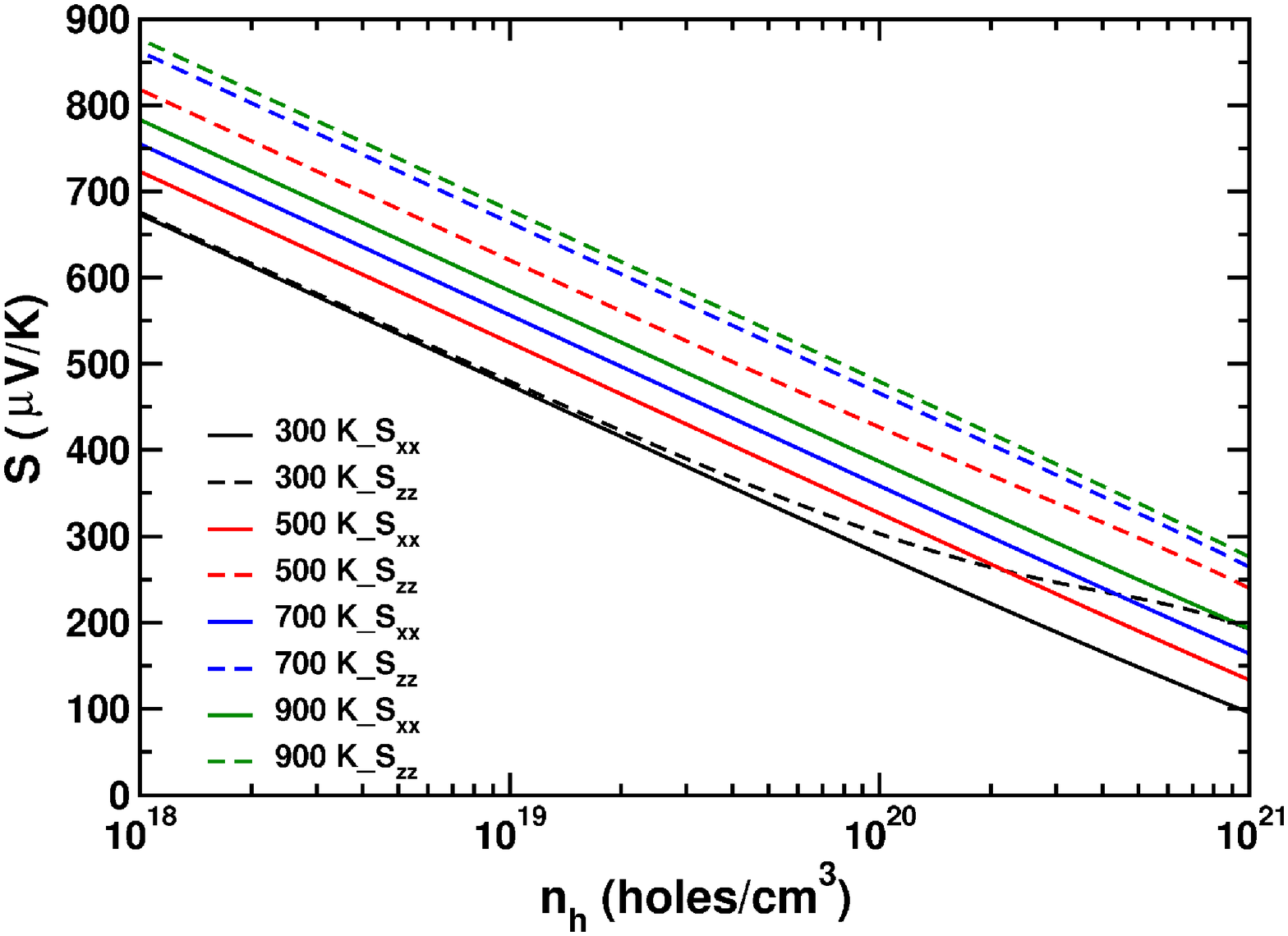}}\\
\subfigure[]{\includegraphics[width=70mm,height=70mm]{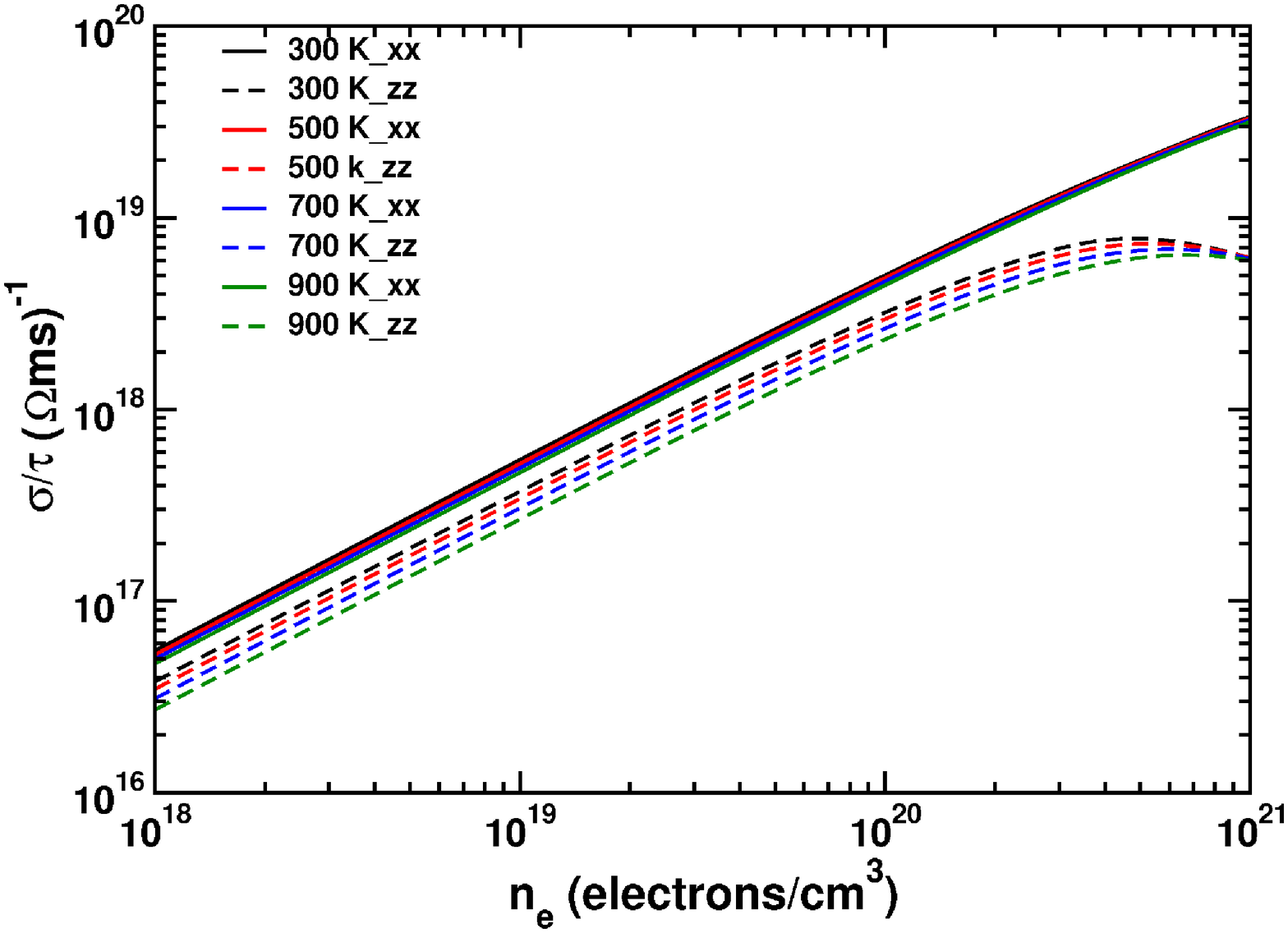}}
\subfigure[]{\includegraphics[width=70mm,height=70mm]{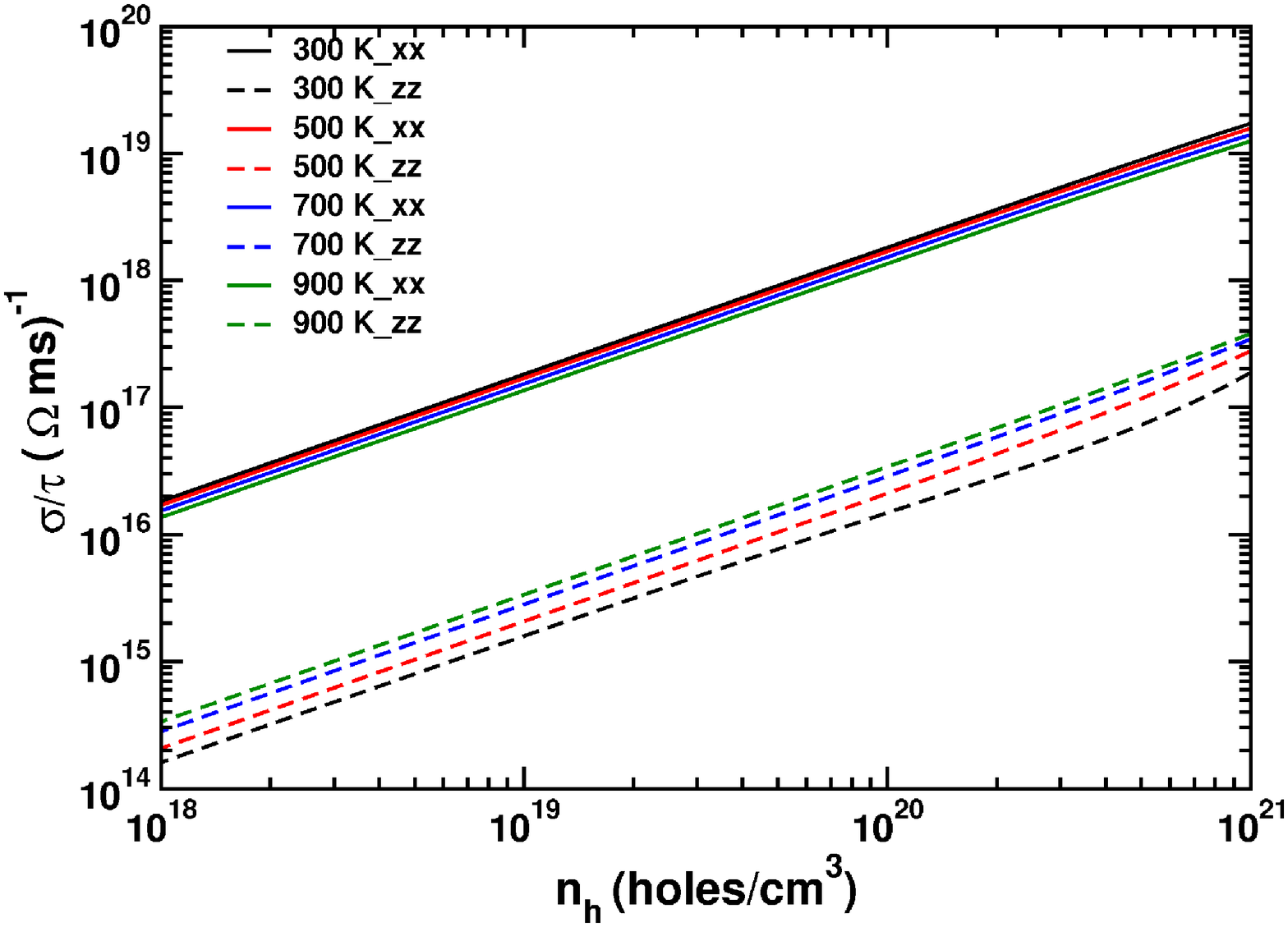}}\\
\caption{(Color online) The calculated thermoelectric properties with variation of carrier concentration at different temperatures of SrAgSF: thermopower for (a): electrons and (b): holes; electrical conductivity scaled by relaxation time for (c): electrons and (d): holes}
\end{center}
\end{figure*}

\begin{figure*}
\begin{center}
\includegraphics[width=70mm,height=70mm]{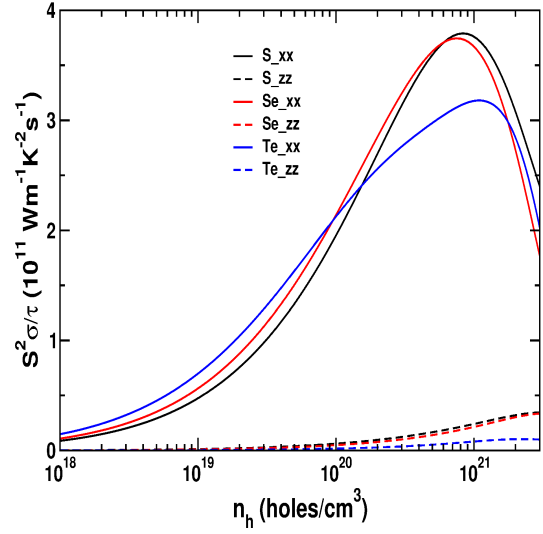}
\caption{(Color online) Calculated power factor variation of SrAg$Ch$F at 700 K}
\end{center}
\end{figure*}

\begin{figure*}
\begin{center}
\subfigure[]{\includegraphics[width=100mm,height=50mm]{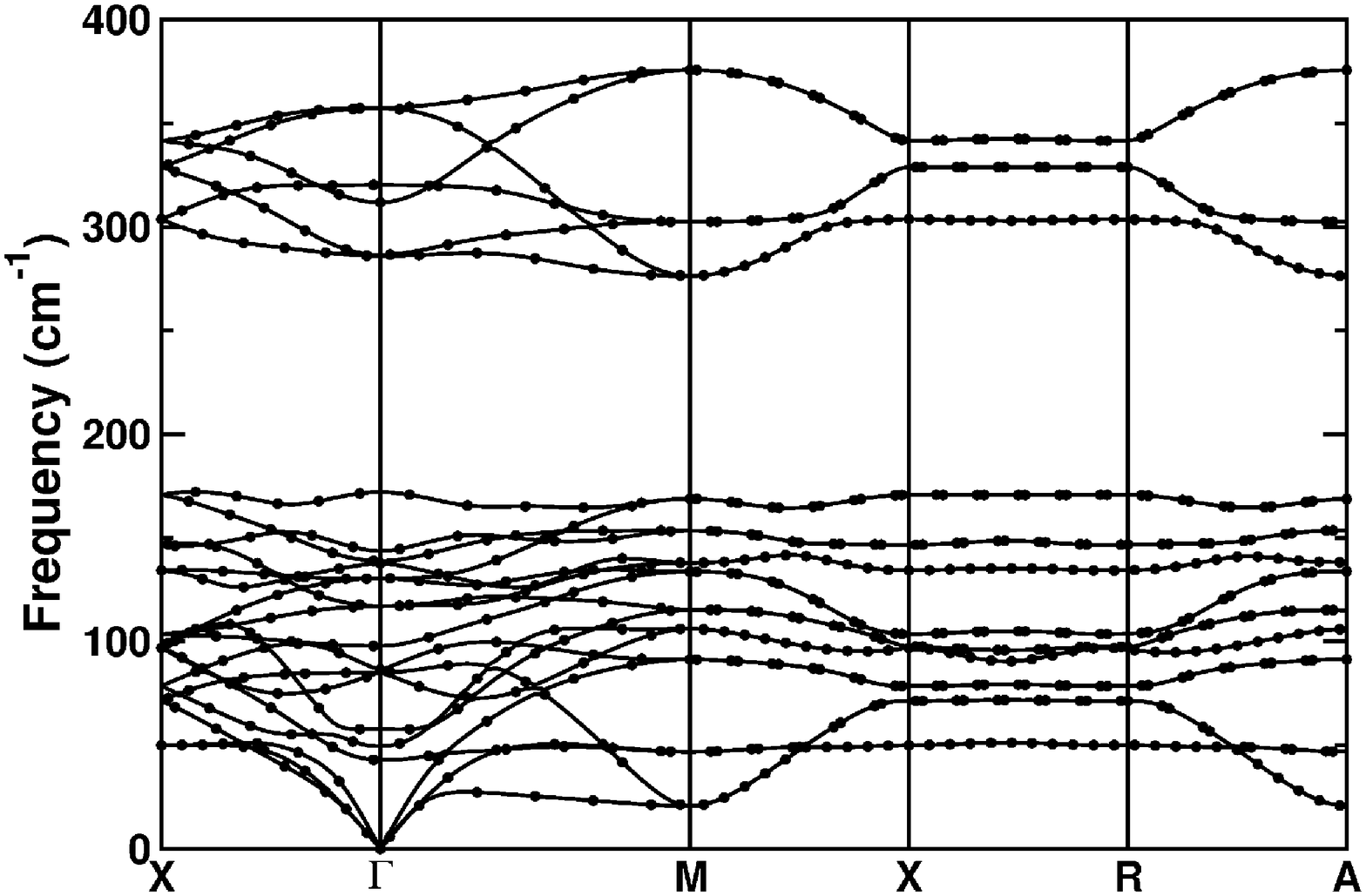}}
\subfigure[]{\includegraphics[width=60mm,height=50mm]{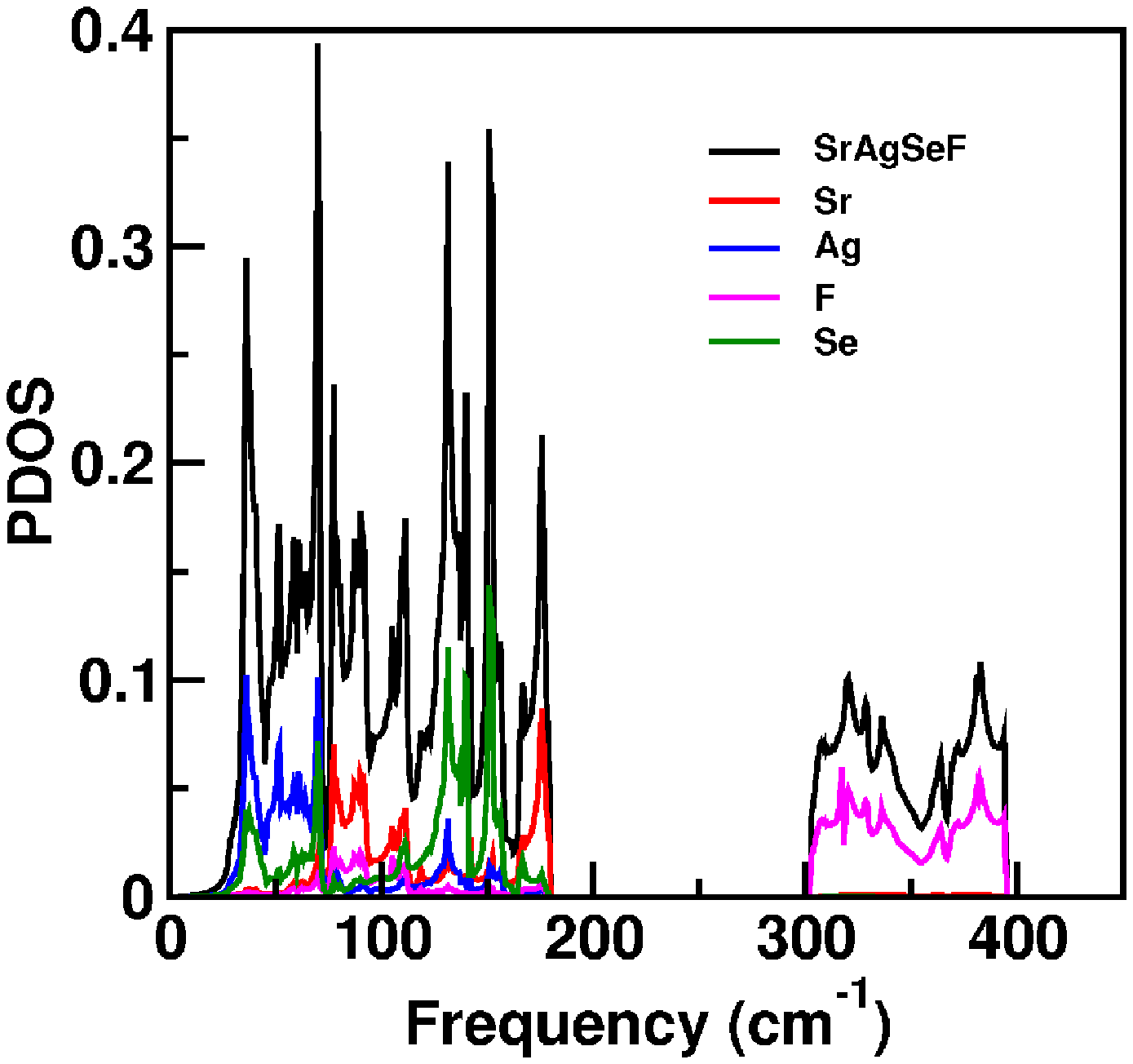}}\\
\subfigure[]{\includegraphics[width=90mm,height=50mm]{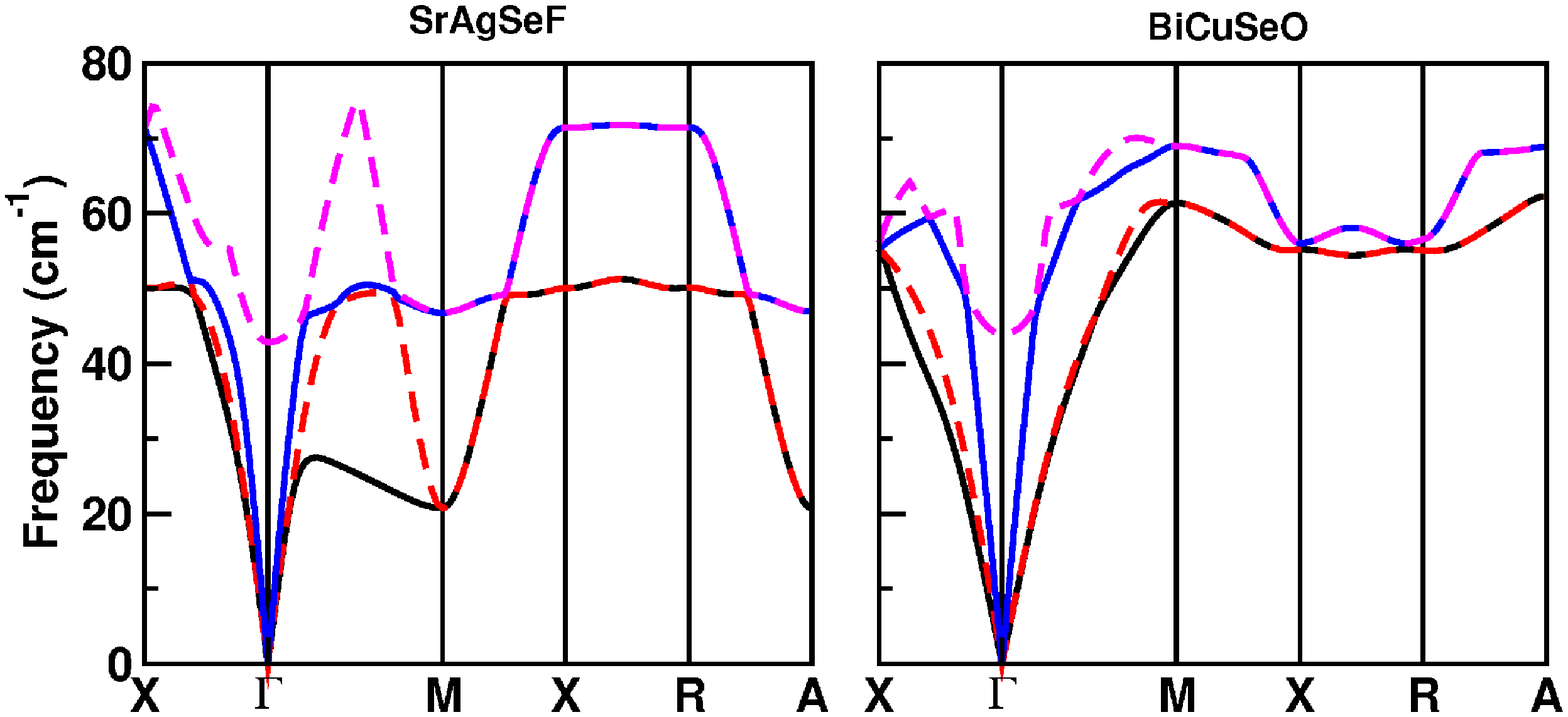}}
\caption{(Color online) (a) Phonon Dispersion curves and (b) partial phonon density of states for SrAgSeF. (c): Comparison of the low-frequency phonon dispersion curves for SrAgSeF and BiCuSeO}
\end{center}
\end{figure*}



\begin{figure*}
\begin{center}
\subfigure[]{\includegraphics[width=65mm,height=65mm]{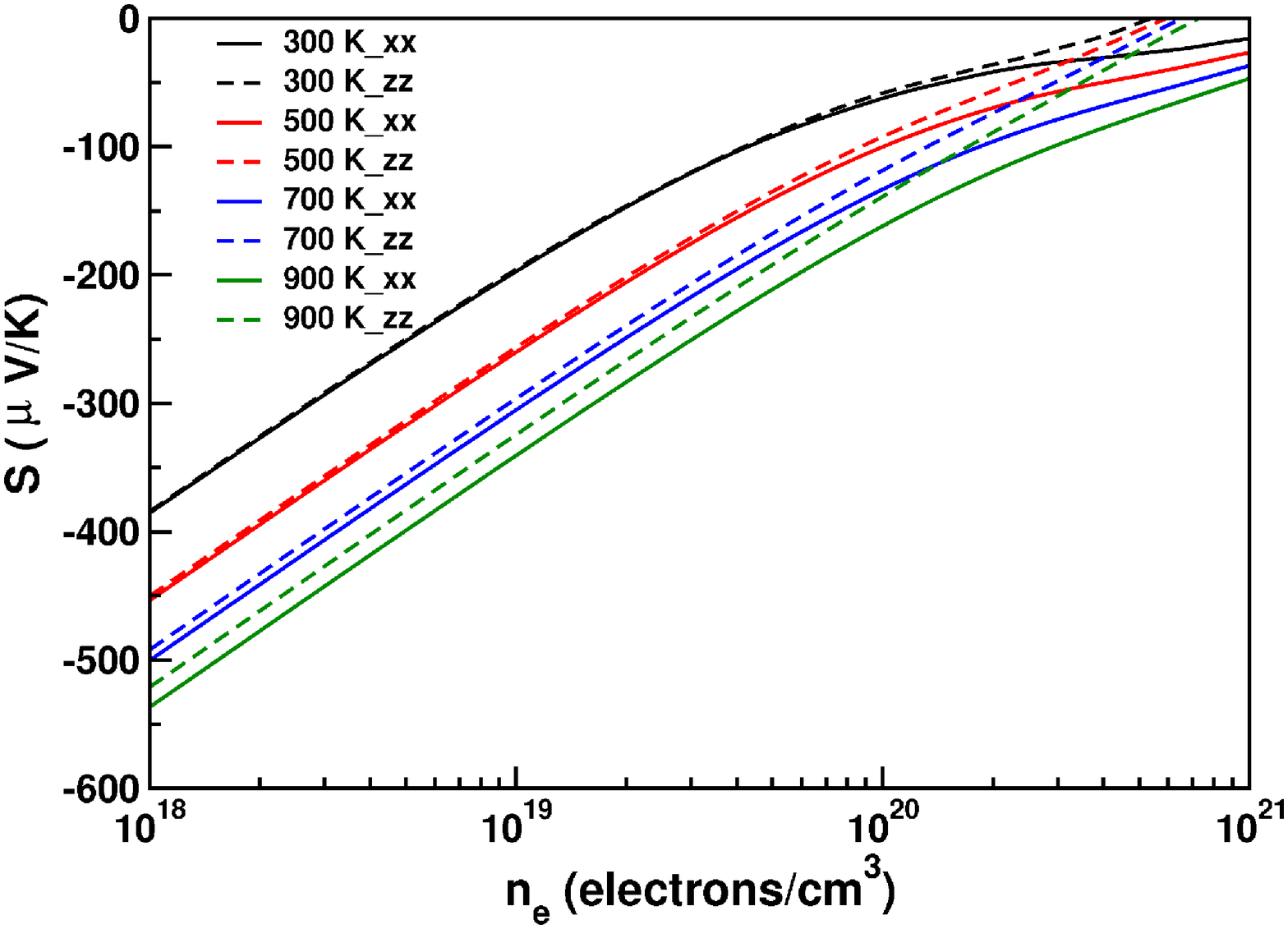}}
\subfigure[]{\includegraphics[width=65mm,height=65mm]{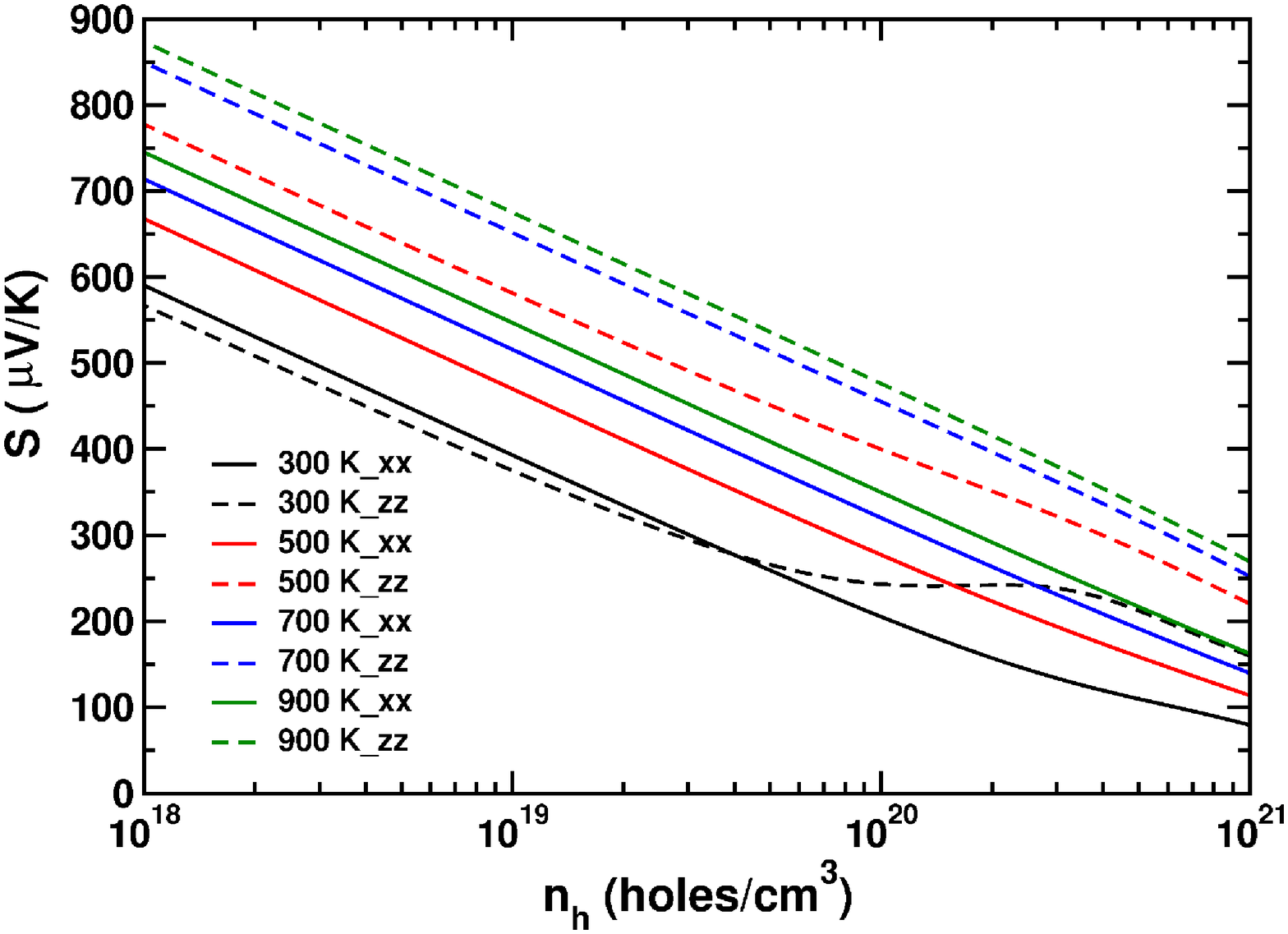}}\\
\subfigure[]{\includegraphics[width=65mm,height=65mm]{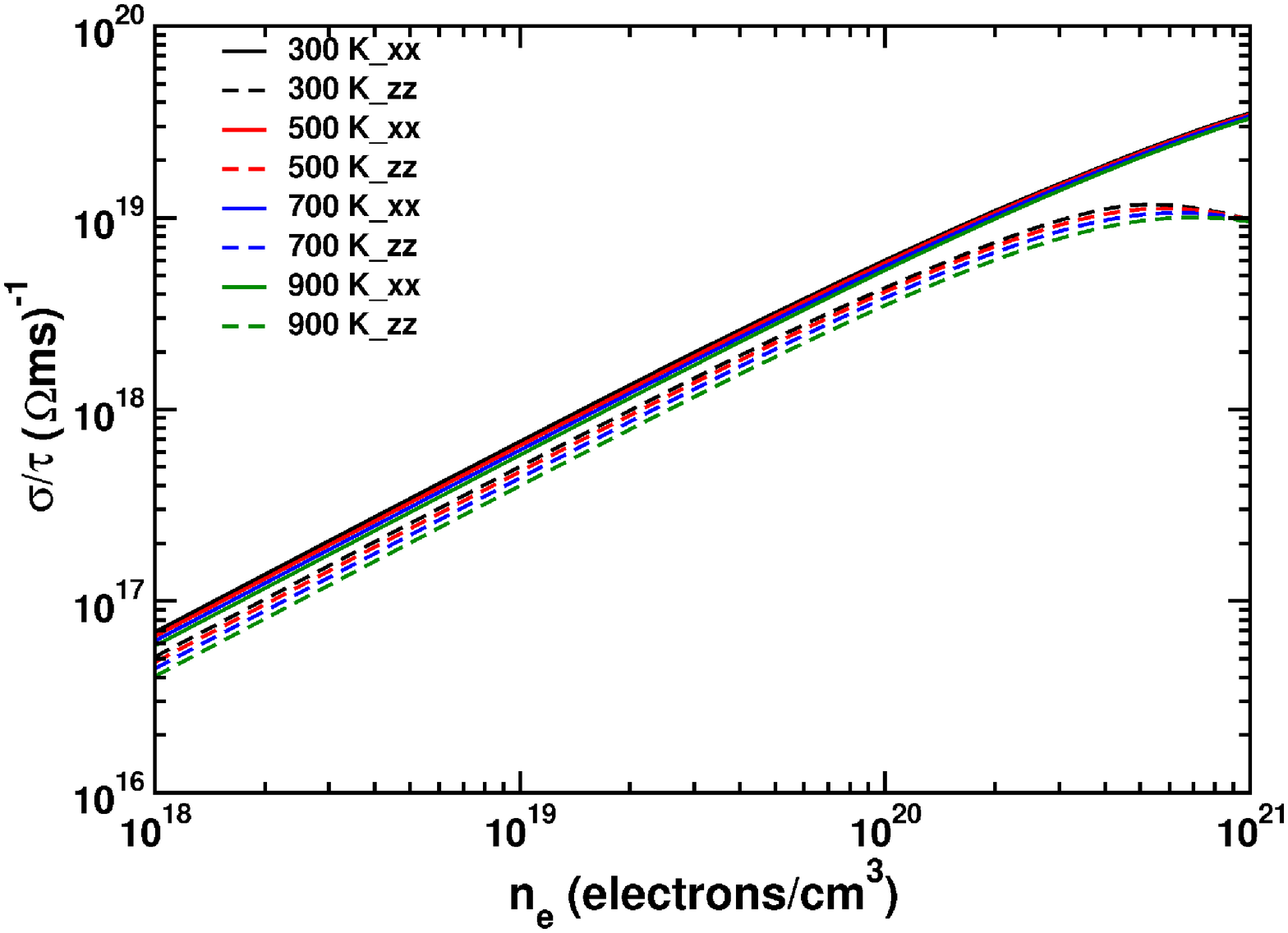}}
\subfigure[]{\includegraphics[width=65mm,height=65mm]{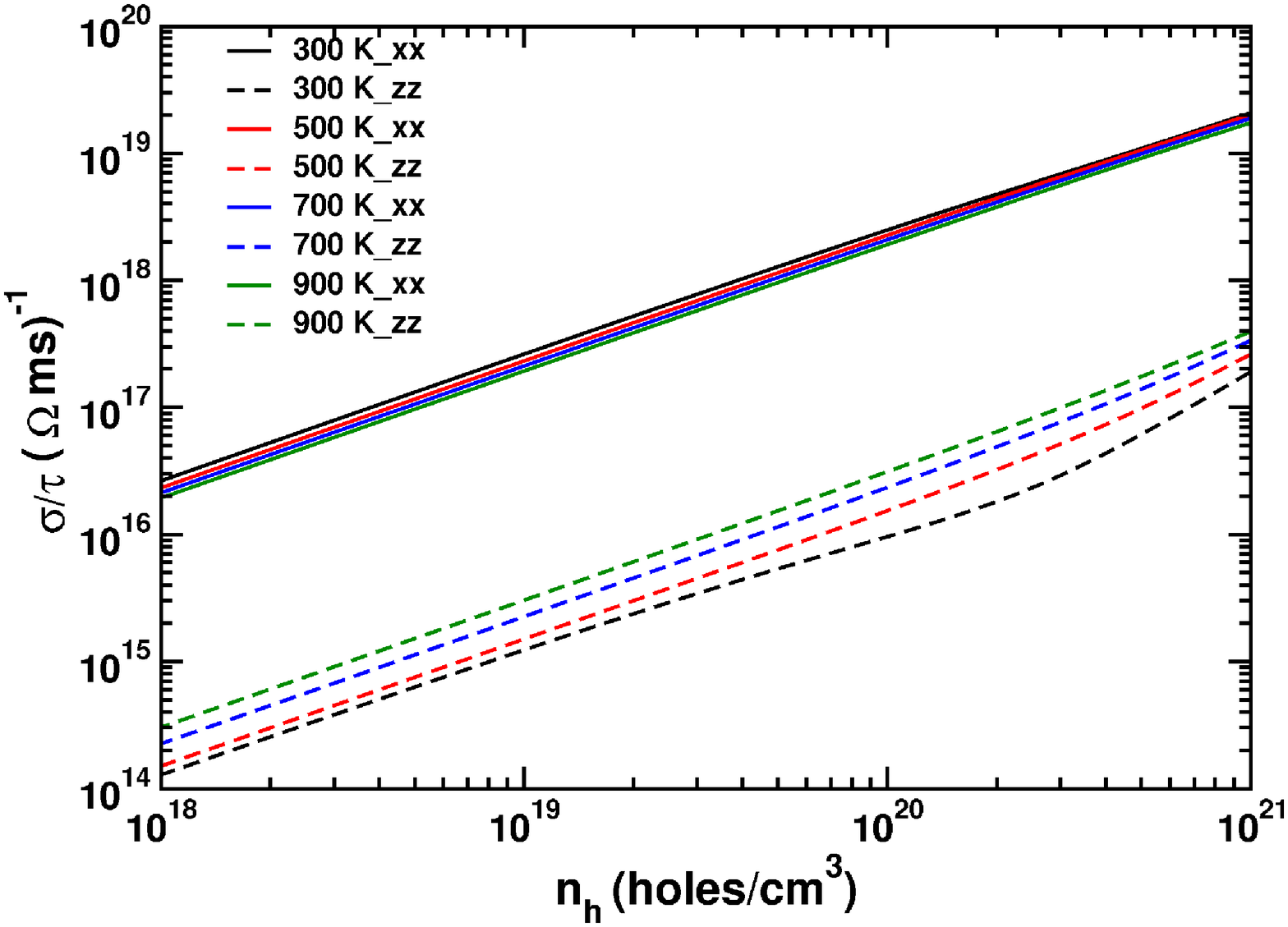}}\\
\caption{(Color online) The calculated thermoelectric properties with variation of carrier concentration at different temperatures of SrAgSeF: thermopower for (a): electrons and (b): holes; electrical conductivity scaled by relaxation time for (c): electrons and (d): holes}
\end{center}
\end{figure*}

\begin{figure*}
\begin{center}
\subfigure[]{\includegraphics[width=70mm,height=70mm]{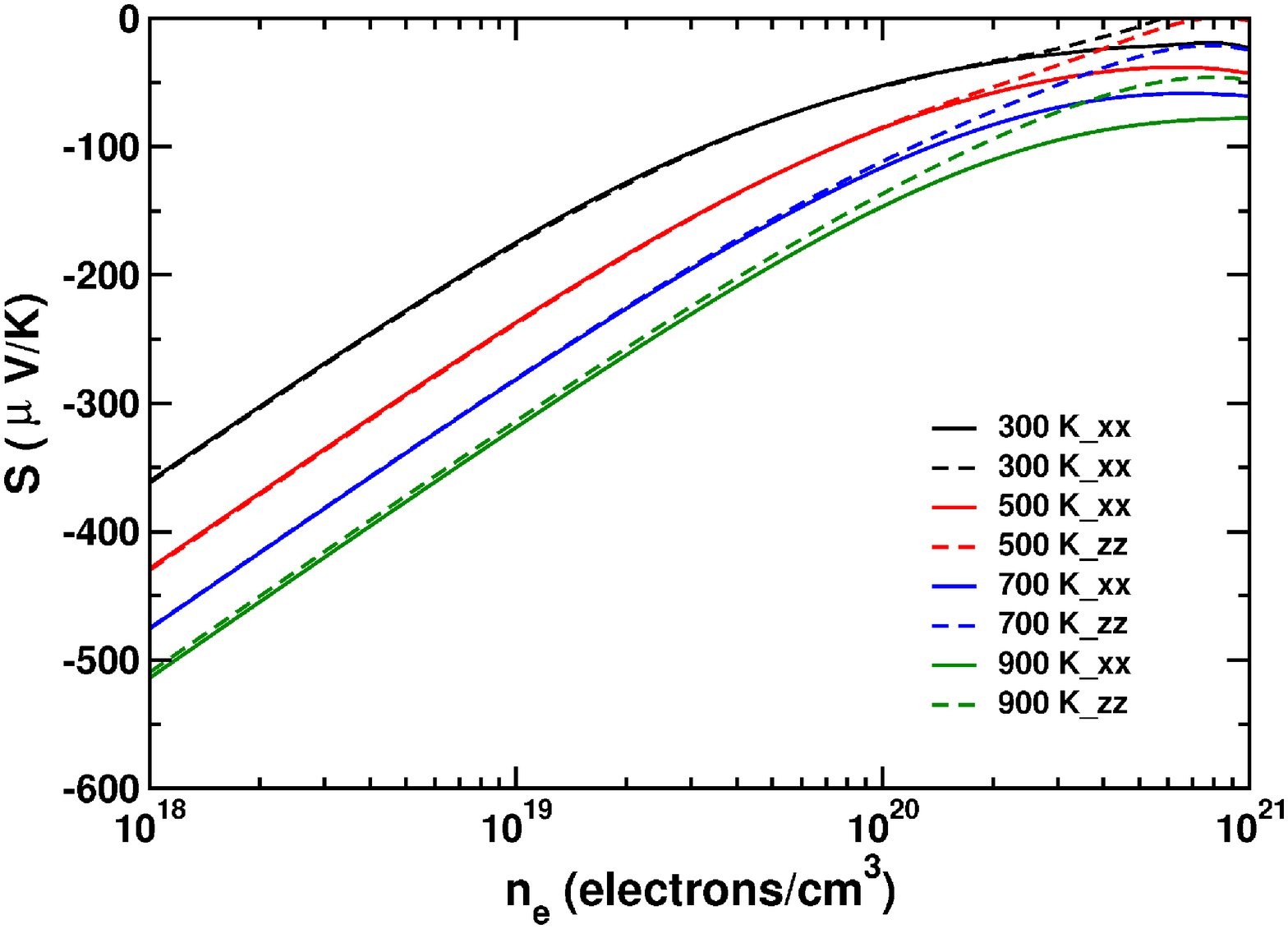}}
\subfigure[]{\includegraphics[width=70mm,height=70mm]{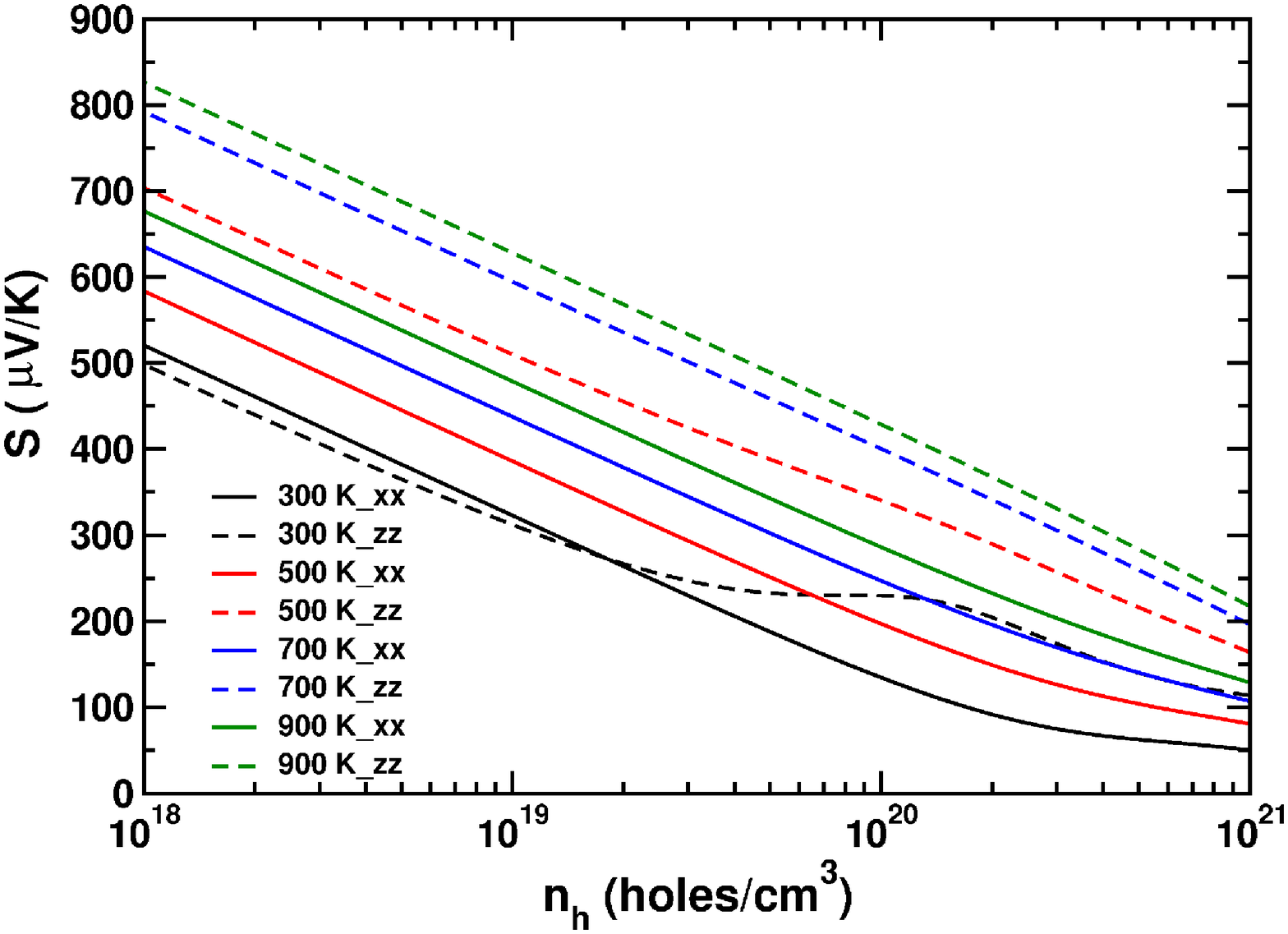}}\\
\subfigure[]{\includegraphics[width=70mm,height=70mm]{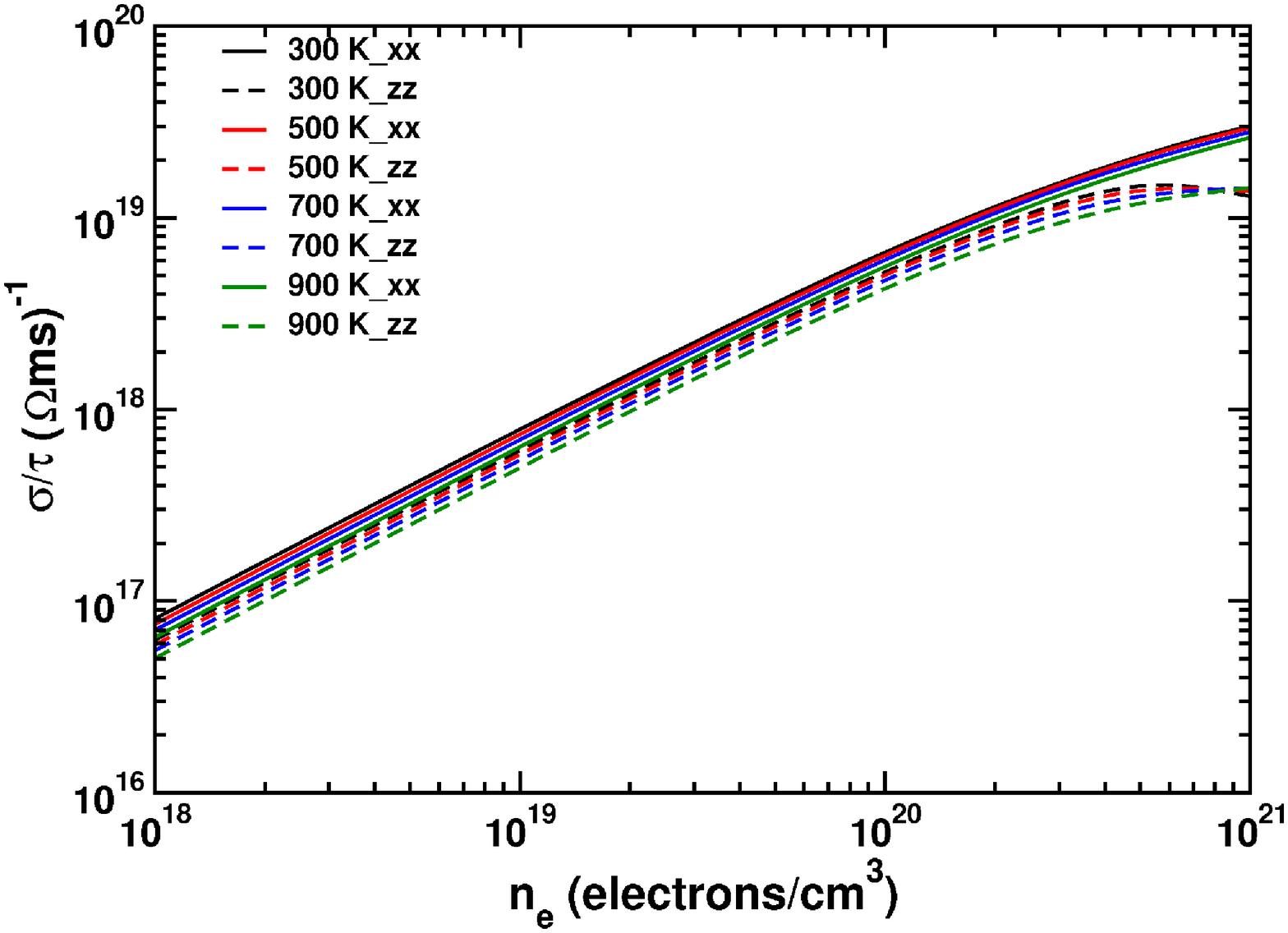}}
\subfigure[]{\includegraphics[width=70mm,height=70mm]{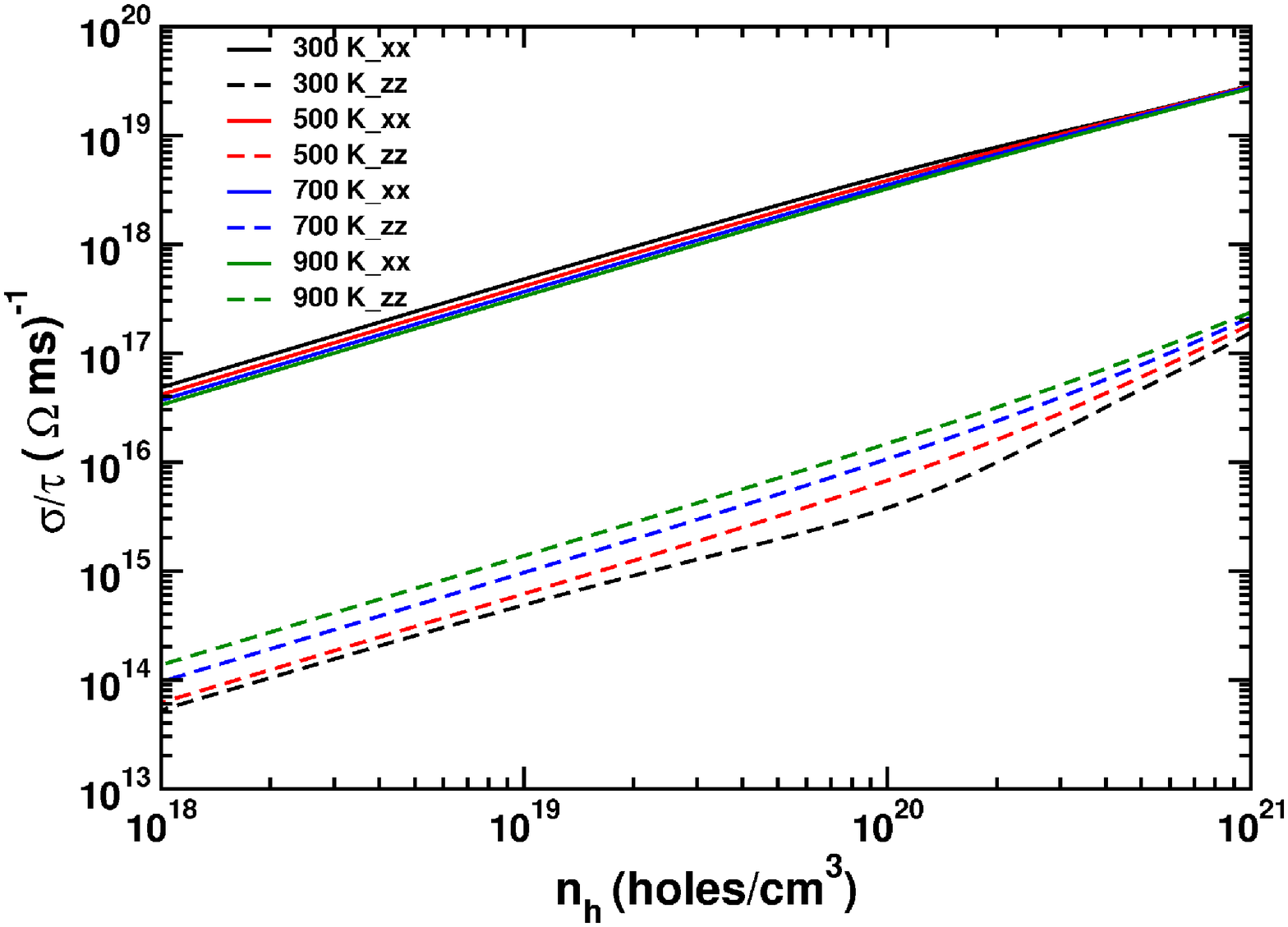}}\\
\caption{(Color online) The calculated thermoelectric properties with variation of carrier concentration at different temperatures of SrAgTeF: thermopower for (a): electrons and (b): holes; electrical conductivity scaled by relaxation time for (c): electrons and (d): holes}
\end{center}
\end{figure*}

\begin{table*}
\caption{Ground state properties of SrAgChF (Ch=S,Se,Te) with GGA functional along with the available experimental and other calculation results (Space Group: 129 (P4/nmm).}
\begin{tabular}{ccccccccccccccccccccccccccc}
\hline
		&SrAgSF		&	&		&SrAgSeF	&	&		&SrAgTeF\\
\hline
		&Present	&Exp	&Other theory &Present	&Exp	&Other theory		&Present	&Exp	&Other theory\\
\hline
a(\AA) 		&4.202 		&4.0593	&4.114		&4.308 		&4.1652	&4.205		&4.421 		&4.3397	&4.368	\\
c(\AA) 		&8.752 		&9.1521	&9.148		&8.895 		&9.2552	&9.367		&9.488 		&9.5947	&9.758	\\
V(\AA$63$) 	&154.58		&150.81	&154.83		&165.14		&160.57	&165.63		&185.51		&180.70	&186.18	\\
		&(+2.5$\%$)	&	&(+2.6$\%$)	&(+2.8$\%$)	&	&(+3.1$\%$)	&(+2.7$\%$)	&	&(+3.0$\%$)\\
z(Sr) 		&0.1565 	&0.1534 &		&0.1506 	&0.1500 &		&0.1374 	&0.1358	\\	
z(Te) 		&0.6921		&0.6945	&		&0.6975		&0.6959	&		&0.6991		&0.6976\\	
Band gap(eV) 		\\
GGA		&1.18		&	&1.25		&1.03		&	&1.09		&1.32		&	&1.33\\
TB-mBJ		&2.53		&	&		&2.27		&	&		&2.33			\\
\hline
Exp:\onlinecite{Charkin}; Other theory:\onlinecite{Bannikov}
\end{tabular}
\end{table*}

\begin{table*}
\caption{Calculated effective mass values of SrAgChF (Ch = S, Se, Te) compared with BaCuChF}
\begin{tabular}{cccccccccccccccccc}
\hline
Effective mass (me)	&SrAgSF		&	&SrAgSeF	&	&SrAgTeF	&	\\
\hline
			&VB		&CB	&VB		&CB	&VB		&CB	\\
\hline
m$_z$			&37.7		&0.96	&33.66		&0.86	&28.79		&2.98	\\
m$_x$			&5.22		&0.69	&5.05		&0.57	&2.87		&0.54	\\
BaCuChF\cite{Zakutayev}	&BaCuSF		&	&BaCuSeF	&	&BaCuTeF\\						
m$_z$			&37.5		&0.92	&34.7		&0.82	&27.0		&2.94	\\
m$_x$			&5.41		&0.68	&4.23		&0.54	&3.85		&0.51	\\
\hline
\end{tabular}
\end{table*}

\begin{table*}
\caption{Atomic and bond population ionicity of SrAgSF:}
\begin{tabular}{cccccccccccccccccc}
\hline
Atomic Populations (Mulliken) 		&&&Bond population ionicity: (covalent $<$ 0.5 $<$ Ionic) \\
\hline
Atom	&Populations 		&&Bond 	&Population\\
\hline
Sr 	&1.24 			&&Sr-F	&0.55(Ionic)\\		
Ag 	&-0.07 			&&Ag-S 	&0.44(Covalent)\\
F 	&-0.62 \\
S 	&-0.55 \\
\hline
\end{tabular}
\end{table*}

\begin{table*}
\caption{Calculated elastic constants of SrAgChF (Ch = S, Se, Te) in comparison with BiCuSO}
\begin{tabular}{cccccccccccccccccc}
\hline
Parameters	&SrAgSF	&	&SrAgSeF	&	&SrAgTeF	&	&BiCuSO\cite{Liu}\\
\hline
		&CASTEP	&WIEN2k	&CASTEP	&WIEN2k	&CASTEP	&WIEN2k		\\
\hline
C$_{11}$ (GPa)	&110.9 	&114.4	&105.7 	&102.7	&87.7	&99.6	&153.0\\
C$_{33}$ (GPa)	&76.4 	&71.1	&44.5 	&61.7	&50.6	&61.0	&118.9\\
C$_{44}$ (GPa)	&31.0 	&45.4	&15.9 	&34.6	&24.9	&31.5	&34.9\\
C$_{66}$ (GPa)	&26.6 	&29.1	&25.5 	&21.7	&18.4	&22.0	&50.2\\
C$_{12}$ (GPa)	&32.8 	&33.5	&40.0 	&30.9	&24.5	&27.0	&63.9\\
C$_{13}$ (GPa)	&41.8 	&40.7	&44.5 	&46.4	&34.4	&42.7	&57.1\\
Bulk Modulus(B) (GPa)	&57.9 &-	&43.2	&-	&43.7	&-	&87.1\\
Anisotropy	&0.79 	&1.06	&0.48	&0.96	&0.79	&0.87	\\
B$_V$(GPa)		&59.0 	&54.7	&57.0 	&57.2	&45.8	&53.9	&86.8\\
G$_V$(GPa)		&29.8 	&39.9	&14.6	&27.7	&22.4	&26.8	&40.4\\
B$_R$ (GPa)	&58.0	&52.7	&43.2	&55.9	&43.9	&52.9	&85.1\\
G$_R$ (GPa)	&28.6 	&37.5	&14.6 	&22.9	&20.0	&23.2	&39.3\\
B$_{VRH}$ (GPa)	&58.5 	&53.7	&50.1 	&56.6	&44.8	&53.4	&-\\
G$_{VRH}$ (GPa)	&29.2 	&38.7	&17.2 	&25.3	&21.2	&24.99	&-\\
E (GPa)		&75.1 	&93.6	&46.4 	&66.1	&55.1	&64.87	&103.6\\
Density (g/cc)	&5.30 	&5.30	&5.9 	&5.9	&6.12	&6.84	&8.31\\
V$_l$(km/s)	&4.29 	&4.45	&3.52 	&3.91	&3.46	&3.56	&4.09\\
V$_t$(km/s)	&2.62 	&2.7	&1.71 	&2.07	&1.86	&1.91	&2.10\\
V$_m$ (km/s)	&2.62 	&2.99	&1.92 	&2.32	&2.08	&2.13	&2.44\\
$\theta_D$ (K)	&290.5 	&331.5	&208.5 	&251.34	&217.1	&231.28	&288.9\\
\hline
\end{tabular}
\end{table*}

\end{document}